\documentclass[11pt]{article}
\usepackage[margin=1in]{geometry}
\usepackage{graphicx}
\usepackage{natbib}
\usepackage{sansmath}
\usepackage{amsthm}
\usepackage{mwe}

\usepackage{floatrow}
\usepackage{blindtext}
\newfloatcommand{capbtabbox}{table}[][\FBwidth]

\newtheorem{example}{Example}

\usepackage[citecolor=blue, linkcolor=blue, colorlinks=true, urlcolor=blue]{hyperref}
\usepackage{amssymb, amsmath, amsthm, mathtools}
\usepackage{color}
\usepackage{graphicx}
\usepackage{setspace}

\newtheorem{theorem}{Theorem}

\newtheorem{remark}{Remark}[section]

\usepackage{enumerate}

\newcommand{\Real}{\mathbb{R}}

\newcommand{\nn}{\nonumber}

\newcommand{\nnn}{n}

\newcommand{\Tn}{T^{\nnn}}
\newcommand{\TTn}{\tilde T^{\nnn}}
\newcommand{\TTnn}{\tilde T^{\nnn'}}

\newcommand{\Yn}{Y^{\nnn}}
\newcommand{\TYn}{\tilde Y^{\nnn}}
\newcommand{\TYnn}{\tilde Y^{\nnn'}}
\newcommand{\Hnp}{H_0^{\mathrm{np}}}

\newcommand{\Yni}{Y^{\nnn,(i)}}
\newcommand{\Ynr}{Y^{\nnn,(r)}}

\newcommand{\yp}{y^{\mathrm{p}}}
\newcommand{\Ypn}{Y^{\nnn,\mathrm{p}}}

\newcommand{\truetheta}{\theta^\ast}
\newcommand{\truepsi}{\psi^\ast}
\newcommand{\eps}{\varepsilon}
\newcommand{\epsn}{\eps^{\nnn}}
\newcommand{\GG}{\mathcal{G}^{\nnn}}
\renewcommand{\gg}{\mathrm{g}}

\newcommand{\Hfull}{H_0^{\mathrm{full}}}
\newcommand{\obs}{\mathrm{obs}}
\newcommand{\pval}{\mathrm{pval}}
\newcommand{\sobs}{s_\obs}
\newcommand{\pr}{\mathrm{pr}}

\newcommand{\EX}{\mathrm{E}}
\newcommand{\thetanui}{\theta^{\mathrm{nuis.}}}

\newcommand{\Unif}{\mathrm{Unif}}
\usepackage{bbm}
\newcommand{\Ind}{\mathbbm{1}}

\renewcommand{\mod}{\mathrm{mod}~}

\usepackage{mathtools}
\newcommand\eqd{\stackrel{\mathclap{\normalfont\mbox{\tiny d}}}{=}}
\newcommand\toP{\stackrel{\mathclap{\normalfont\mbox{\tiny p}}}{\to}}

\newcommand{\ThmOne}{
Suppose that Assumptions \eqref{A1}~and~\eqref{A2} hold. Then, the $p$-value in~\eqref{eq:pval_full} is exact in 
finite samples under $\Hfull$, that is, 
for any finite $n>0$,
$$
\pr\left\{\pval(\theta_0, \psi_0) \le \alpha \mid \Hfull\right\} = \alpha.
$$
}

\newcommand{\ThmTwo}{
Suppose that Assumptions \eqref{A1}~and~\eqref{A2} hold. Then, $\Theta_{1-\alpha}$ is a finite-sample valid 
$100(1-\alpha)\%$ confidence set for $\truetheta$; i.e., for any finite $n>0$,
$$
\pr(\truetheta\in\Theta_{1-\alpha} \mid H_0) \ge 1-\alpha.
$$
}

\newcommand{\ThmThree}{
Suppose that Assumptions \eqref{A1}~and~\eqref{A2} hold, and 
that $\hat\psi_{\theta_0} \toP \truepsi$ under $H_0$. Then, $\hat\Theta_{1-\alpha}$ is an asymptotically valid 
$100(1-\alpha)\%$ confidence set for $\truetheta$; i.e.,  as $n$ increases
$$
\pr(\truetheta\in\hat\Theta_{1-\alpha} \mid H_0) \ge 1-\alpha + o_P(1).
$$
}

\newcommand{\ThmFour}{
Suppose that Assumptions \eqref{A1} and~\eqref{A2} hold with $\GG=\Pi(\Tn; \theta_0)$. Then, the $p$-value from Procedure~2 is exact under $\Hnp$ conditionally on the observation times, that is, 
$$
\pr\left\{\pval(\theta_0) \le \alpha \mid \Hnp, \Tn \right\} = \alpha.
$$
It follows that the $p$-value is also exact unconditionally.
}

\newcommand{\MakeProcedure}[2]{
\vspace{5px}
\begin{center}
\textsc{Procedure #1}
\end{center}
\singlespacing
\vspace{-20px}
#2
\onehalfspacing
}

\title{Randomization Inference of Periodicity in Unequally Spaced Time Series with Application to Exoplanet Detection
}

\author{
Panos Toulis \\
University of Chicago \\
Booth School of Business \\
\url{ptoulis@chicagobooth.edu}
\and
Jacob Bean \\
University of Chicago \\
Department of Astronomy and Astrophysics \\
\url{jbean@astro.uchicago.edu}
}

\date{}

\begin{document}
\maketitle

\begin{abstract}
The estimation of periodicity is a fundamental task in many scientific areas of study. Existing methods rely on 
theoretical assumptions that the observation times have equal or i.i.d. spacings, and that common estimators, such as the periodogram peak, are consistent and asymptotically normal.
In practice, however, these assumptions are unrealistic as observation times usually exhibit deterministic  patterns ---e.g., the nightly observation cycle in astronomy--- that imprint
nuisance periodicities in the data. These nuisance signals also affect the finite-sample distribution of estimators, which can substantially deviate from normality. Here, we propose a set identification method, fusing ideas from randomization inference and partial identification. In particular, we develop a sharp test for any periodicity value, and then invert the test to build a confidence set.
This approach is appropriate here because the construction of confidence sets does not rely on assumptions of regular or well-behaved asymptotics. 
Notably, our inference is valid in finite samples when our method is fully implemented, 
while it can be asymptotically valid under an approximate implementation designed to ease computation. 
Empirically, we validate our method in exoplanet detection using radial velocity data. In this context, our method correctly identifies the periodicity of the confirmed exoplanets in our sample. For some other, yet unconfirmed detections, we show that the statistical evidence is weak, which illustrates the failure of traditional statistical techniques.
Last but not least, our method offers a constructive way to resolve these identification issues via improved observation designs. In exoplanet detection, these designs suggest meaningful improvements in identifying periodicity even when a moderate amount of randomization is introduced in scheduling radial velocity measurements.
\end{abstract}

\onehalfspacing


\newpage
\section{Introduction}\label{sec:intro}

The identification of periodicity in time series data is a fundamental task across scientific fields. 
Applications include the classical study of periodic mass extinctions~\citep{raup1986biological}, 
gene expression cycles~\citep{cho1998genome}, circadian rythms in biology~\citep{costa2013inference}, 
and seasonal oscillations in epidemiology~\citep{murray2011seasonal}.
Typically, observations times in these applications are unequally spaced, which 
introduces identification issues that standard statistical methods, be it frequentist or Bayesian, cannot handle.
When the hidden true periodicity cannot be point-identified, it is an open question how statistical inference should proceed, 
and whether it is possible to anticipate such identification issues through better observation designs.

Our work in this paper is motivated by the analysis of radial velocity data in exoplanet detection.
The idea behind this method is to look for periodic variations in velocity data from a star.
When these variations cannot be explained by known signals (e.g., rotation or magnetic activity of the star), they are attributed to an (unseen) exoplanet orbiting around the star. In recent years, the deployment of 
an extensive network of high-resolution, Earth-based observatories has led to astonishing 
exoplanet detections, including ``51 Pegasi b"~\citep{mayor1995jupiter} ---awarded the 2019 Nobel Prize in Physics--- and even a potential discovery in our immediate stellar neighborhood of  Proxima Centauri~\citep{anglada2016terrestrial}. 

The statistical methodology behind these exoplanet detections proceeds in two main steps. 
The first step is to test for any sign of periodicity. If this test succeeds, the second step is to estimate the hidden periodicity.
 %The first step is to test whether there is any periodicity in the observed data. 
To test periodicity, statistical methods rely on the so-called {\em periodogram} of the data, which essentially quantifies the power of the Fourier spectrum of the data across different periods~\citep{schuster1898investigation}.
For instance, the classical method of~\citet{fisher1929tests} relies on a statistic 
defined as the ratio of the periodogram peak over its average value across all periods, which has a 
known distribution under a harmonic model with i.i.d.~normal errors.
This method has since been improved in terms of power~\citep{siegel1980testing, bolviken1983new, chiu1989detecting}, and  has been extended to more general hypotheses~\citep{juditsky2015detecting} 
and sparse alternatives~\citep{cai2016global}.
For adaptations of these tests to astronomical applications see also
\citep{baluev2008assessing, baluev2013detecting, delisle2020efficient, linnell1985test}, 
and the concept of ``false alarm probability".
In general, testing for periodicity presents no big challenges as current tests 
are robust and their properties are well-understood.

However, the problem of actually estimating the hidden periodicity is 
more subtle. For one, standard parametric results assume away identification issues, typically by assuming that the true periodicity lies in a constrained space allowing point-identification with an increasing number of observations~\citep{walker1971estimation, hannan1971non, hannan1973estimation, quinn1991fast, quinn2001estimation}; this is also related to the problem of aliasing and 
the Nyquist limit in statistical Fourier analysis~\citep[Section 2.5]{bloomfield2004fourier}.
Furthermore, standard methods that rely on Fourier analysis also require equally spaced observation times. In practice, however, observations are usually unequally spaced for practical reasons, 
such as the difficulty in gathering fossils to study biological extinctions or the logistics of scheduling 
astronomical observations with large telescopes.

To address the  problem of unequally spaced observation times,~\citet{lomb1976least} and \citet{scargle1982studies} independently developed a modification of Schuster's periodogram. 
Even though the Lomb--Scragle periodogram was motivated by Fourier analysis, 
it has an intuitive statistical interpretation as  it is  equivalent to  fitting the data with a harmonic model through least squares. As such, the Lomb--Scragle periodogram currently occupies a central place in exoplanet detection methodology~\citep[Section 1.1]{vanderplas2018understanding}, where the focus is on assessing whether the periodogram peak is statistically significant, 
typically via extreme value theory. 

A common flaw in current methodology, however, is that a significance result on the periodogram peak is interpreted to mean that the true unknown period is also ``somewhere near" the peak.
The implicit underlying assumption is, of course, that the finite-sample distribution of the periodogram peak is 
near the true period, and also normal in shape. This is unrealistic, however, 
because the likelihood in the harmonic models employed in practice is highly non-smooth and multimodal with respect to the underlying period. As such, without strong, unrealistic assumptions on the observation times, the periodogram peak is not ``well-behaved" neither in finite samples nor asymptotically, in the sense that it may deviate significantly from normality, and may even be inconsistent.
See also~\citep{hall2006using} for similar discussions on the challenges of estimating periodicity through periodograms.
Technically, one fundamental problem is that the separability conditions required for consistency of M-estimators~\citep[Theorem 5.7]{van2000asymptotic}, such as the peak of a periodogram, generally do not hold under arbitrary observation designs.  
Below, we present some examples to illustrate these issues and motivate our work.

\begin{example}[Synthetic data]
\label{example1}\em
Let  $t_i=i + 0.05 U_i$, $i=1, \ldots, 100$, be the observation times for data
$y_i = 1.5\cos(2\pi t_i/\sqrt{2}) + \eps_i$, where $U_i\sim \Unif[-1, 1]$ and $\eps_i\sim N(0, 1)$ i.i.d. The true periodicity is $\truetheta=\sqrt{2}\approx 1.414$.
The data and its generalized Lomb--Scragle periodogram~\citep{zechmeister2009generalised} are shown in Figure~\ref{fig0}~(left). In our sample, the peak is at $\hat\theta_n = 3.417$ and is  distinct 
from the peak around the true period. There are also multiple peaks 
associated with the true period harmonics. Using Fisher's test we can reject 
the null of no periodicity~($p$-value is 0).

\begin{figure}[t!]
\centering
\includegraphics[scale=0.36]{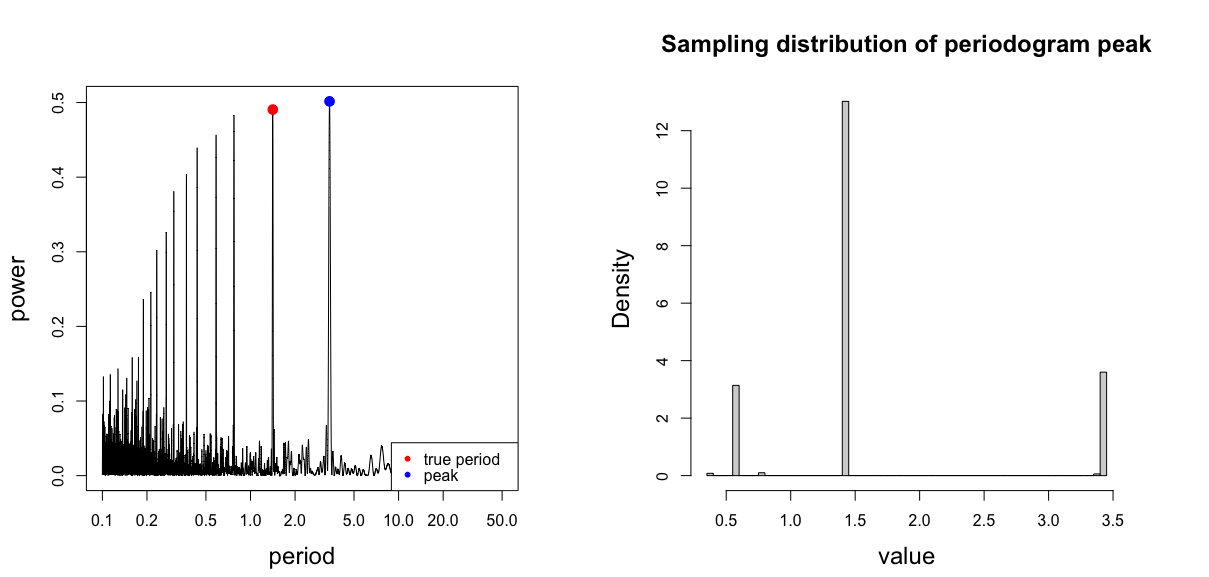}
\caption{
\small
{\em Left:} Periodogram from one dataset from model of Example~\ref{example1}. The 
dataset was seeded to illustrate the identification issue.~{\em Right:} Sampling distribution of the periodogram peak from the same model over 1,000 replications. We see that 
the periodogram peak is not normal, which complicates inference.}
\label{fig0}
\end{figure}

Inference on $\truetheta$, however, is challenging as we need to know the sampling distribution of $\hat\theta_n$. 
%The tacit assumption, in practice, is that $\hat\theta_n$ is normal around $\theta_0$, so that estimation can follow standard statistical techniques. 
As argued above, normality of $\hat\theta_n$ ---as it is commonly assumed in practice--- is not plausible without  strong assumptions on the observation times, especially in finite samples.
Indeed, in this example, we see that the simulated sampling distribution of $\hat\theta_n$ is far from normal, and instead looks discrete with three distinct modes, as shown in Figure~\ref{fig0}~(right). The empirical density 
over 1,000 simulations is shown below~(rounded to two significant digits):
\renewcommand{\arraystretch}{1.2}
\begin{table}[h!]
\begin{tabular}{c | lllllllll}
$\hat\theta_n$   & 0.37  & 0.59   & 0.77  & 1.41   & 1.42   & 3.40   & 3.41  & 3.42  & 3.43  \\
\hline
$\pr(\hat\theta_n | \theta_0)$ & 0.4\% & 15.3\% & 0.2\% & 55.6\% & 10.9\% & 1.6\% & 7.6\% & 7.8\% & 0.6\%
\end{tabular}
\end{table}
\end{example}

We see that the sampling distribution of our estimator is non-regular. 
This complicates inference as it is unclear, for example, how to construct confidence intervals for the true period. In this paper, we propose a method that can transform such sampling distributions into 
valid confidence sets that can still cover the true period with a target coverage probability.

\begin{example}[Extinction cycles]
\label{example2}
\em
In a series of influential articles,~\citet{raup1986periodic} and \citet{raup1986biological} analyzed 
data from biological extinction events, and attempted to estimate their periodicity.
Their approach did not use the periodogram, but it modeled the event times at the most prominent extinction events directly for a best-matching period. Significance was determined through a resampling scheme where all the  recorded extinction times (not just the prominent ones) were permuted.~Through this method,~\citet{raup1986periodic} determined that there exists a biological extinction cycle of roughly 26 million years.
However, this result may be an artifact of the small sample size (only 40 observations), and the unequal spacing of observation times.~Indeed, \citet{stigler1987substantial} showed that the same signal at 26 million years appears even in simulated data where 
the observation times are fixed but the actual data are random.
We analyze the extinction data in Section~\ref{sec:extinction} using the method proposed in this paper, and present a statistically principled complement to the heuristic analysis in~\citep{stigler1987substantial}.
\end{example}

\begin{example}[Planet around~$\alpha$ Centauri B?]
\label{example3}
\em
In a widely-publicized article,~\citet{dumusque2012earth} reported the detection of an Earth-mass planet around 
the star $a$ Centauri B at a period of 3.24 days. This detection was startling because $\alpha$ Centauri is the closest stellar system to our own system, implying 
that habitable exoplanets may be ubiquitous. However, the detection has not yet been confirmed, 
and there have been some early objections to it~\citep{hatzes2013radial}. Notably,~\citet{rajpaul2015ghost} followed a methodology similar to~\citep{stigler1987substantial} discussed in~Example~\ref{example2}, 
and showed that the 3.24-day signal also appears as a local peak in various simulations with quasi-random data and fixed observation times. In this paper, we analyze these data in Section~\ref{sec:exoplanet} and show that the statistical evidence is weak in the sense that the observed data could be explained by many other candidate signals. Moreover, in Section~\ref{sec:design} we 
discuss how the observation design could be improved to identify the 
3.24-day signal, assuming that the signal is true.
\end{example}

We emphasize that neither bootstrap- nor Bayesian-based methods have so far addressed this type of identification issues. 
On the one hand, bootstrap procedures require random observation times and convergence in distribution for $\hat\theta_n$, which requires strong, unrealistic assumptions on the design. Moreover, the small-sample performance of bootstrap procedures with time series is theoretically not well understood. This is particularly important in exoplanet detection as we usually have to work with only a few hundred observations.
Bayesian methods, on the other hand, have the problem of prior specification. 
For example, even through they are usually referred to as ``non-informative", uniform priors 
give preference to parameter regions that not only have high likelihood but  are also wide. 
In effect,  this sweeps the identification problem ``under the rug" as narrow parameter regions that cause identification problems~(such as the region around $\theta =0.59$ in Example~\ref{example1}) are washed away; see also~\citep[Section 1]{hall2003nonparametric} for similar arguments that invoke number theory, and~\citep[Section 3]{gelman2020holes} for a more general discussion on how 
uniform priors can unintentionally lead to sharp  inference.
In a Bayesian context, this issue also spills over to model selection through Bayes factors in the sense that 
 model decisions based on Bayes factors may strongly depend on  features that are esoteric to the selected models~(e.g., size or structure of the parameter space); see also~\citep[Section 6]{gelman2020holes} for details.
 
\subsection{Our contribution}
In this paper, we propose a set identification method to infer hidden periodicity. 
The main idea is to develop a sharp test for a null hypothesis on a given periodicity value, 
$H_0: \theta_0 = \theta$, and then invert the test to build a confidence set, $\Theta_{1-\alpha}$. Similar to a confidence interval, a confidence set 
is constructed so that it contains the true parameter with a pre-specified coverage probability:
\begin{align}\label{eq:thetan}
\pr(\truetheta \in \Theta_{1-\alpha} ) \ge 1-\alpha.
\end{align}
%Our method works with both equally or unequally spaced time series, and can also work, in principle, for any choice of statistics other than periodogram peaks.
In contrast to confidence intervals, confidence sets are not  necessarily contiguous.
The use of confidence sets is therefore more suitable here since 
it does not require asymptotic normality, or even consistency, of periodogram peaks.
Furthermore, the ``topology" of $\Theta_{1-\alpha}$ can reveal identification issues based on, say, 
how wide or narrow the confidence set is.

We propose two constructions for $\Theta_{1-\alpha}$. 
The first construction is idealized and yields a finite-sample valid procedure, where~\eqref{eq:thetan} holds exactly in the sample~(Theorem~\ref{thm1}). 
This construction is computationally challenging, in practice, and so we also propose an alternative  construction, where~\eqref{eq:thetan} holds asymptotically~(Theorem~\ref{thm3}). This approximate construction may be conservative  but in empirical applications we show this not to be a major concern.
Our approach borrows ideas from randomization inference in statistics~\citep{lehmann2006testing, rosenbaum2010design}, where inference is  based on structural assumptions (e.g., symmetries) on the data, and is therefore robust~\citep{toulis2019life}. Our approach is also related to partial identification in econometrics~\citep{manski2003partial, romano2008inference, tamer2010partial}, where test inversion is used to build confidence sets for unknown parameters that cannot be point-identified.

While several heuristic techniques have been used to diagnose identification issues in periodicity estimation~(see Examples~\ref{example1} and~\ref{example2}), 
our paper is the first to develop a valid, principled statistical procedure for inference when such issues are present. 
The proposed procedure is described in detail in Section~\ref{sec:concrete}, and can in fact work with many different choices of the statistic (other than periodogram peaks), and can be parallelized for computational efficiency.
%
% The confidence set, $\Theta_{1-\alpha}$, produced from our procedure not only reveals any identification issues, but also quantifies the severity of the problem.
Moreover, our procedure does not require normality or i.i.d.~errors, 
and can seamlessly work with equally or unequally spaced data.
We note that while some important non-parametric methods with unequally spaced data have recently been proposed, they still require observation designs with equal spacings~\citep{sun2012nonparametric, vogt2014nonparametric} or
i.i.d.~spacings between observation times~\citep{hall2000nonparametric, hall2008nonparametric, hall2003nonparametric}, and emphasize 
asymptotic consistency over finite-sample performance.
%  is unrealistic in practice as ground-based observations typically need to happen at night and during the summer, and thus exhibit strong deterministic patterns.
%
That said, our procedure could leverage these methods in the choice of the test statistic.
This could combine the best of both worlds, namely the finite-sample validity and robustness from our randomization-based procedure, and the flexibility and power from nonparametric methods.
See~Remark~\ref{remark:statistic} and Section~\ref{sec:np} for more discussion.

Empirically, we illustrate our method in several examples from exoplanet detection using radial velocity data.
In this context, we show that our method gives sharp inference on the confirmed exoplanets in our sample, namely 51 Pegasi b~\citep[``51Pegb"]{mayor1995jupiter} and Gliese 436 b~\citep[``GJ436b"]{butler2004neptune}, showing that our inference is not conservative.
For other, yet unconfirmed, exoplanets in our sample our method gives more mixed results.
For example, we show that the statistical evidence is notably weak for a candidate discovery 
around $\alpha$ Centauri B~\citep{dumusque2012earth} in the sense that the confidence set, $\Theta_{1-\alpha}$, contains a large range of signals, some possibly non-planetary, including nuisance 1-day signals that are usually associated with the nightly cycle of astronomical observations. 
Testing another widely-publicized discovery in Proxima Centauri~\citep{anglada2016terrestrial}, 
we find that the discovery seems to be robust apart from a nuisance signal at 0.916 days that cannot be rejected at the 1\% level.
 Importantly, our method suggests ways to improve the observation designs for sharper inference. Specifically, we quantify how much additional randomness is needed in the observation times or, alternatively, how many more observations are needed to precisely identify a candidate signal that is suspected to be true.
 Following this approach, we propose designs that could address the aforementioned identification issues, and also help with future scientific discoveries.

\section{Problem Setup and Notation}\label{sec:setup}
We observe data $\Yn = \{ y_1, \ldots, y_n\}\in\mathbb{R}^n$ at observation times  
$\Tn = \{t_1, t_2, \ldots, t_n\}$. For example, in exoplanet detection, the data
could represent measured radial velocity of a star at a time $t$ expressed as a barycentric Julian date.
% In contrast to previous literature, we only assume that the observation times are independent of the signal:
%\begin{align}\label{A1}
%\Tn \indep \Yn.\tag{A1}
%\end{align}
%This assumption is weak as  it is reasonable to expect that stellar observations are unaffected by the time of observation. In contrast, earlier work makes stronger assumptions about the design, requiring either equally spaced data together with strong identification conditions, or i.i.d.~spacings between observations.
%
We decompose the data model as $y_i = \yp(t_i) + \eps(t_i)$, where $\yp(t)$ is an 
unobserved periodic, non-stochastic function with unknown period $\truetheta$, such that 
$$
\yp(t+\truetheta) = \yp(t),~\quad(t\in\Real).
$$
The terms $\eps(t_i)$ correspond to the unobserved errors.  We use 
$\Ypn=(\yp(t_1), \ldots, \yp(t_n))$ and $\epsn = (\eps(t_1), \ldots, \eps(t_n))$ 
to denote the (unobserved) vector values of these two components.

Our main goal is to do inference on $\truetheta$, the true unknown period. 
We  will need the following additional assumptions.
First, we require a consistent estimator of $\yp$, which could either be 
parametric  or nonparametric.
For simplicity, in the following section we start with the simple but popular trigonometric regression model for $\yp$. 
Later, in Section~\ref{sec:np}, we also discuss a nonparametric approach.
Second, we assume that the errors, $\eps(t)$, form a stochastic process satisfying the following 
invariance property. For any observation times $\Tn = \{t_1, \ldots, t_n\}$,  with $n$ finite, there exists 
a known algebraic group $\GG$ of $n\times n$ matrices such that 
\begin{align}\label{A1}
\gg\cdot \epsn\eqd \epsn~\mid \Tn,~\quad(\gg\in\GG).\tag{A1}
\end{align}

This formulation follows the framework of randomization tests~\citep[Chapter 15]{lehmann2006testing}, 
where testing is based on structural assumptions rather than analytical.
To keep things simple, for the rest of this paper we assume that $\GG = \mathbb{D}^n$, the set of $n\times n$ diagonal matrices with $\pm 1$ in the diagonal. 
This definition implies that the regression errors, $\eps(t)$, are assumed to have a sign symmetric distribution.
As such, our inference method not only covers Gaussian errors, which are frequently assumed in practice, 
but also extends to more general model classes where
the error distribution is unspecified, or even varies with $t$, as long as the symmetry in~\eqref{A1} holds.
%
% As such, our inference method not only covers Gaussian errors, which are frequently assumed in practice, 
% but also extends to model classes with general error distributions.

% The following remark adds some more details on the choice of $\GG$.
% Moreover, Assumption~\eqref{A2} can also accommodate more complex errors; see Remark~\ref{rem21} below for more details.

%The following remark provides some more discussion on the generality of~\eqref{A3}.

\begin{remark}[Generality of~\eqref{A1}]
\label{rem21}
\em
Assumption~\eqref{A1} can also allow inference with complex error structures.
%Working with invariance assumptions rather than distributional assumptions allows us to perform inference under general conditions on the errors.
%
% As such, with this definition of $\GG$, our method can work with any symmetric distribution of the errors  beyond Gaussian.
Suppose, for instance, that $\eps(t)$ is a
 Gaussian process~(GP) with covariance matrix $C$. 
Then, we could define $\GG = \{ \gg' = C^{1/2} \gg C^{-1/2} : ~\gg\in\mathbb{D}^{n}\}$.
Of course, this requires $C$ to be known, or be consistently estimated.
To accomplish that, one idea would be to combine our approach with 
recent GP-based methods on radial velocity data~\citep{rajpaul2015gaussian, damasso2017proxima,faria2016uncovering,faria2020decoding}. We leave this for future work.~$\blacksquare$
\end{remark}

\section{Main Method}\label{sec:method}
Here, we follow a parametric approach to model the periodic component, $\yp$. 
A nonparametric approach is discussed later, in Section~\ref{sec:np}.
All proofs can be found in the Supplement.

We start with the simple harmonic model 
\begin{align}\label{A2}
\yp(t) \equiv \yp(t; \truetheta, \truepsi) = \psi^\ast_{1} + \psi^\ast_{2} \cos(2\pi t/\truetheta) + \psi^\ast_{3} 
\sin(2\pi t/\truetheta),
\tag{A2}
\end{align}
where $\truetheta\in\Theta\subseteq\mathbb{R}^+$ is the true 
unknown period, and $\truepsi = (\truepsi_1, \truepsi_2, \truepsi_3) \in \Psi\subseteq\mathbb{R}^3$ are unknown nuisance parameters, 
where $\Psi$ is convex and compact. We note that we use this model for simplicity. 
For instance,  we could work with, say, a Keplerian model that includes additional parameters to describe various patterns of motion (e.g., eccentricity) of a body relative to another~\citep{wright2009efficient}.
All that is required for our theory to work is for the model of choice to be consistently estimated conditional on 
the true period, $\truetheta$.

To measure the fit of $\yp$ we use the simple squared loss,
$$
L(\theta, \psi \mid \Yn, \Tn) = \sum_{i=1}^n [y_i - y^{\mathrm{p}}(t_i | \theta, \psi)]^2 / \sigma_i^2,
$$
where the precisions $\sigma_i^2$ are known from instrument specifications, as is typical in exoplanet detection.
Note that the loss function~(i.e., negative log-likelihood under Gaussian errors) is non-convex and non-smooth with respect to $\theta$, and so standard statistical inference cannot be applied.

To proceed, let $L_{0n}(\Yn, \Tn) = \sum_{i=1}^n (y_i - \bar y)^2/\sigma_i^2$ denote the baseline loss, where 
$\bar y=(1/n) \sum_{i=1}^n y_i$.
 Then, the {\em oracle} generalized periodogram, $A_n: \Theta\to\mathbb{R}^+$ can be defined as follows:
\begin{align}\label{eq:oracle_per}
A_n(\theta \mid \Yn, \Tn, \psi) = \frac{L_{0n}(\Yn, \Tn)  - L(\theta, \psi \mid \Yn, \Tn) } { L_{0n}(\Yn, \Tn)  }.
\end{align}
In practice, $\psi$ is not the target of inference, and so we use the profile likelihood
\begin{align}\label{eq:real_per}
\hat A_n(\theta \mid \Yn, \Tn) \equiv A_n(\theta \mid \Yn, \Tn, \hat\psi_\theta),
~\text{where}~\hat\psi_\theta = \arg\min_{\psi\in\Psi} L(\theta, \psi \mid \Yn, \Tn).
\end{align}
$\hat A_n$ is known as the generalized periodogram, and is easy to calculate as $\hat\psi_\theta$ is known in closed form~\citep{zechmeister2009generalised}. Our first goal is to test a global null hypothesis of the form:
\begin{align}\label{eq:H01}
\Hfull: \truetheta=\theta_0, \truepsi=\psi_0.
\end{align}
This hypothesis is not particularly interesting because, ultimately, we are only interested in the unknown period $\truetheta$. However, testing for $\Hfull$ will serve as a good starting point for testing and inference on $\truetheta$.
Of course, to test any hypothesis we need a test statistic, $s_n(\Yn, \Tn)\in\mathbb{R}$. 
Many choices are possible here, and so we defer a detailed discussion about this choice to Section~\ref{sec:concrete}.
Here, we simply define
% Fisher's classical statistic for testing periodicity~\citep{fisher1929tests} and define
$$
 s_n(\Yn, \Tn) = \max_{\theta \in \Theta} A_n(\theta \mid \Yn, \Tn, \psi_0) - A_n(\theta_0 \mid \Yn, \Tn, \psi_0).
%s_n(\Yn, \Tn) = |\Theta| \max_{\theta \in \Theta} A_n(\theta \mid \Yn, \Tn, \psi_0) / \sum_{\theta\in\Theta} A_n(\theta \mid \Yn, \Tn, \psi_0).
$$
The idea behind this choice of the test statistic is that $s_n$ will tend to take small values when $\truetheta=\theta_0$, as we expect the periodogram to peak around $\theta_0$ under $\Hfull$; 
when $\Hfull$ is not true we expect $s_n$ to take large values as the periodogram will peak in 
periods different from $\theta_0$, so that the null can be rejected. 
% This choice also relates to the classical Fisher statistic for testing periodicity~\citep{fisher1929tests}, but empirically it leads to sharper inference.

Let $\sobs = s_n(\Yn, \Tn)$ denote the observed value of the test statistic in the sample. 
To construct the null distribution of $s_n$, we simulate data as follows:
$$
\Yni = \Ypn_0+ G^{(i)} \cdot (\Yn - \Ypn_0),
$$
where $\Ypn_0 = [\yp(t_1; \theta_0, \psi_0), \ldots, \yp(t_n; \theta_0, \psi_0)]$ are the periodic component values, which are known under the null hypothesis; $G^{(i)}\sim\mathrm{Unif}(\GG)$ is a random error transformation from $\GG$. 

Then, a  $p$-value for $H_0^{\mathrm{full}}$ is
\begin{align}\label{eq:pval_full}
\pval(\theta_0, \psi_0) = E\{s_n(\Yni, \Tn) \ge \sobs \mid  \Yn, \Tn\},
\end{align}
where the randomness is with respect to $G^{(i)}$ while $\Yn, \Tn$ are fixed. 
The following result shows that this $p$-value is valid in finite samples for testing $\Hfull$.

\begin{theorem}\label{thm1}
\ThmOne
\end{theorem}

%% Nuisance parameters.
In most settings, $\truepsi$ is a nuisance parameter, and so we only want to test for $\truetheta$:
\begin{align}\label{eq:H02}
H_0: \truetheta=\theta_0.
\end{align}
In this context, the result in Theorem~\ref{thm1} can still be used, in principle, to build a confidence set for 
$\truetheta$ using the following construction: 
\begin{align}\label{eq:Theta1}
\Theta_{1-\alpha} = \left\{\theta \in \Theta :   \max_{\psi\in\Psi}  \pval(\theta, \psi) > \alpha \right\}.
\end{align}

\begin{theorem}\label{thm2}
\ThmTwo
\end{theorem}
%\begin{proof}
%Suppose that the null hypothesis $H_0$ with $\theta_0, \psi_0$ is true. Then, 
%$v(\theta_0, \psi_0) \sim U[0,1]$.
%\begin{align}
%P(\theta_0\in\Theta_{1-\alpha} \mid D_{\by}, \bt)
% & = P\{\max_{\psi} v(\theta_0, \psi) \ge \alpha \mid D_{\by}, \bt\} \nonumber\\
% & \ge  P\{v(\theta_0, \psi_0) \ge \alpha \mid  D_{\by}, \bt\}
%= P\{U \ge \alpha\} = 1-\alpha.
%\end{align}
%\end{proof}
While $\Theta_{1-\alpha}$ is a valid confidence set, its construction is challenging due to the optimization problem 
in~\eqref{eq:Theta1} that requires maximizing the $p$-value over the entire nuisance parameter space, $\Psi$. This is especially challenging when $\Psi$ is multidimensional. 
To address this computational issue, a feasible procedure is  to estimate 
$\truepsi$ under the null hypothesis~$H_0$ ---for example, through the least-squares estimate $\hat\psi_{\theta_0}$--- and then plug  this estimate into~\eqref{eq:Theta1}. 
We note that this approach works because the estimator $\hat\psi_{\theta_0}$ is consistent for $\truepsi$  under the standard assumptions of least squares since, under fixed $\theta$, the loss function in the harmonic model is smooth and convex with respect to $\Psi$.
In contrast, estimating $\truetheta$ in a similar fashion (e.g., through least squares) does not work, 
precisely because of the identification issues discussed in Section~\ref{sec:intro}.

% \todo{LS conditions?}

To test $H_0$ in a computationally manageable way, we therefore have to simply re-define the test components using the generalized periodogram~\eqref{eq:real_per} as follows:
\begin{align} 
 s_n(\Yn, \Tn) &  = \max_{\theta \in \Theta} \hat A_n(\theta \mid \Yn, \Tn) - \hat A_n(\theta_0 \mid \Yn, \Tn),
 \label{eq:re1} \\
 \hat Y^{\mathrm{p}}_0 &  = [\yp(t_1; \theta_0, \hat\psi_{\theta_0}), \ldots, \yp(t_n; \theta_0,\hat\psi_{\theta_0})], \label{eq:re2}\\
 \Yni  & = \hat Y^{\mathrm{p}}_0 + G^{(i)} \cdot [\Yn - \hat Y^{\mathrm{p}}_0 ]. \label{eq:re3}.
\end{align}
The $p$-value for $H_0$ is then defined as
\begin{align}\label{eq:pval_theta}
\widehat\pval(\theta_0) = E\{s_n(\Yni, \Tn) \ge \sobs \mid \Yn, \Tn\},
\end{align}
where, again, the randomness is with respect to $G^{(i)}$ while the data $(\Tn, \Yn)$ are 
fixed.
The following construction for the confidence set of $\truetheta$ is valid asymptotically:
\begin{align}\label{Theta2}
\hat\Theta_{1-\alpha} = \left\{\theta \in \Theta :  \widehat\pval(\theta) > \alpha \right\}.
\end{align}

\begin{theorem}\label{thm3}
\ThmThree
\end{theorem}

% \todo{What if $e(t)$ have approximate symmetry? Leverage result by Canay et al..}

\subsection{Concrete procedure and practical considerations}\label{sec:concrete}
In summary, we propose the following procedure to estimate the true periodicity.

\vspace{-5px}
\MakeProcedure{1}{
\begin{enumerate}
\item Choose a grid of possible period values, $\Theta$ that contains $\truetheta$ w.p. 1.
Set $\hat\Theta_{1-\alpha} \gets \emptyset$.
\item Obtain data $(\Yn, \Tn)$, possibly after removing known 
stellar signals, e.g., rotational periods, magnetic cycles, etc.~\citep[Section 11.3.4]{feigelson2012modern}.
\item {\bf For} all $\theta_0\in\Theta$ {\bf do:}
\begin{enumerate}[(i)]
\item Estimate the nuisance parameters, $\hat\psi_{\theta_0} = \arg\min_{\psi\in\Psi} L(\theta_0, \psi \mid \Yn, \Tn)$, through weighted least squares.
\item Calculate the observed value, $\sobs$, of the test statistic 
defined in~\eqref{eq:re1}.
\item With fixed $\Tn$, sample new data, $\Yni$, where $i=1, \ldots, R$ for some fixed $R$, according 
to definitions~\eqref{eq:re2} and \eqref{eq:re3}.
\item Using the samples  from 3(iii), calculate the $p$-value~\eqref{eq:pval_theta}, and 
if it exceeds $\alpha$ then include $\theta_0$ in the confidence set;
i.e., set $\hat\Theta_{1-\alpha} \gets \hat\Theta_{1-\alpha} \cup \{\theta_0\}$ 
if $\widehat\pval(\theta_0) > \alpha$.
\end{enumerate}
\item {\bf Return} $\hat\Theta_{1-\alpha}$ as the $100(1-\alpha)\%$ confidence set of $\truetheta$.
\end{enumerate}
}

In words, Procedure~1 considers all possible candidate period signals in $\Theta$. 
For every candidate signal, $\theta_0$, the procedure, first, estimates the harmonic model assuming that the true period is $\theta_0$, and then 
it randomly flips the signs of the residuals to generate new synthetic data~(i.e., assumes symmetric distribution for the errors).
The observed value of the test statistic is then compared against the test statistic values in the synthetic data, and the resulting $p$-value determines whether $\theta_0$ should be in the confidence set or not.

We end this section with a few remarks on some practical aspects of implementing Procedure~1.
\begin{remark}[Choice of $\Theta$]
\label{remark:Theta}
\em
The choice of $\Theta$ can be crucial as the periodogram is generally sensitive to the average spacing in $\Theta$. In general, it is a good idea to choose spacings that are smaller than the expected width of the major periodogram peaks, and also induce a spacing in $\Theta$ that is uniform in the log scale~\citep[Section 6.1]{vanderplas2018understanding}.
Another useful technique is to keep refining $\Theta$ until the shape of the observed periodogram and its major peaks are stable.~$\blacksquare$
\end{remark}

\begin{remark}[Choice of test statistic, $s_n$]
\label{remark:statistic}
\em
Procedure 1 is valid for any choice of the test statistic. Our choice of test statistic in~\eqref{eq:re1} performs well under several controlled simulations~(see Section~\ref{sec:example}), but other options are certainly possible. For instance, Fisher's classical statistic to test periodicity 
can be defined as $s_n = \max_{\theta \in \Theta} \hat A_n(\theta) 
/ \bar A_n$, where $\bar A_n = |\Theta|^{-1}\sum_\theta A_n(\theta)$ is the average periodogram value; here, we suppressed the dependence on $\Yn, \Tn$ to ease notation. Improvements using a trimmed mean in place of 
$\bar A_n$ have also been suggested~\citep{bolviken1983new, siegel1980testing, damsleth1982estimation}, including sparse alternatives~\citep{cai2016global}. See also~\citep{mcsweeney2006comparison} for some numerical comparisons.~$\blacksquare$
\end{remark}

\begin{remark}[Power]
\label{remark:power}
\em
The power of Procedure 1 depends on how sensitive the chosen test statistic, $s_n$, is in detecting violations of the null hypothesis. In general, statistics that lead to better estimators of $\truetheta$ will also lead to more powerful tests, and thus more narrow confidence sets. In this paper, we don't explore the power of our randomization tests theoretically to maintain the focus on our application---we refer the reader to~\citep{toulis2019life} and~\citep{dobriban2021consistency} for such theoretical investigations on the 
power of randomization tests, including those applied in a regression context like ours.
Some optimal results in testing periodicity are known but require strong assumptions on the errors. 
For instance, Fisher's classical statistic (described above) is optimal 
only under i.i.d. normal errors in the harmonic model. Empirically, our choice of test statistic in~\eqref{eq:re1} appears to be powerful compared to alternatives---see Section~\ref{sec:example} and Section~\ref{sec:exoplanet} for details.~$\blacksquare$
\end{remark}

%\begin{remark}[Structured errors]
%\em 
%In this paper, we have assumed that the errors $\eps(t)$ have a symmetric distribution.
%Our method can accommodate additional structures, however, with an appropriate definition of $\GG$. For instance, if $\eps(t_i)$ are not symmetric but are exchangeable, we could define $\GG$ to be the set of $n\times n$ permutation matrices. Correlated structures could also be addressed as described in Remark~\ref{rem21}. See also~\citep{toulis2019life} about the use of randomization tests with even more complex error structures~(e.g., clustering or autocorrelation).
%\end{remark}

\begin{remark}[Computation]
\label{remark:computation}
\em 
The complexity of Procedure~1 can be described, prima facie, as $O(|\Theta|^2 \cdot R \cdot C)$, 
where $C$ is the average computation time for a single weighted least-squares fit. 
For some ballpark numbers, we can take $|\Theta| = 10^4$, $R=10^3$, and $C=50\mu$s for an analysis on a conventional laptop of a time series with 200 observation times, using an implementation in {\tt R}.
The full computation requires  a total of 1,388 hrs.~of wall clock time~(approx. 58 days).
However, several reductions of computation time are possible. 
First, Procedure 1 can be fully parallelized since step 3 can be executed independently for each $\theta_0\in\Theta$. With a moderate-sized cluster of 100 computing nodes, the 
wall clock time thus drops to 14 hrs. Moreover, again in step 3, there is no need to consider all values in $\Theta$ but only a proportion of the periodogram peaks. In our applications, for example, 
we only consider those peaks of the periodogram that are at least 20\% as high as the highest peak.
This leads to a complexity $O(\gamma |\Theta|^2 \cdot R \cdot C)$ with $\gamma$ ranging between
$0.1\%$-$3\%$ in our applications.
As such, the required computation drops dramatically to roughly 30 mins.~of wall clock time.
Indeed, in our application, we are able to ramp up our samples to $R=100,000$ and still 
finish all exoplanet analyses within a couple of hours using a cluster of 400 computing nodes.
Further significant gains are possible if the method is implemented in a more efficient language than {\tt R}, such as 
{\tt C++} or {\tt Python}.~$\blacksquare$
\end{remark}

\section{Numerical Examples}\label{sec:example}

\subsection{Extinction data}\label{sec:extinction}
Here, we return to Example~\ref{example2} on estimating periodicity in biological extinction events.
The original data are shown in the left subplot of Figure~\ref{fig:extinction} below.
% comprising of 40 observations of extinction events over time measured as \% of estimated biomass lost during the event.
~\citet{raup1986periodic} argued that there is a 26 million year periodicity based on fitting 
a simple periodic model on the largest extinction events. Formally, for any given $\theta\in\Theta$, they fit a model $t_i = t_0 + k_\theta \theta$ for those timepoints that correspond to the eight largest peaks; here $t_0, k_\theta$ are free parameters and $k_\theta$ is an integer.
Using this model,~\citet{raup1986periodic} calculated which $\theta$ yields the smallest least squares fit, and then used a permutation-based test to assess significance.

The periodogram of the data is shown in the right subplot of Figure~\ref{fig:extinction}.
Indeed, using a set $\Theta$ ranging between 12.5-62.5 million years, as suggested by~\citet{raup1986periodic}, we find the largest periodogram peak to be around 26 million years.
However, we also see additional peaks around 30 million years and 35 million years.
Using Procedure~1, none of these values can be rejected as the true periodicity~($p$-values are 0.56 and 0.6, respectively).\footnote{All data and code for the numerical examples of this section, including the exoplanet detection applications of the next sections, are  provided with the Supplementary Code. 
An online repository with {\tt R} code to replicate all results and figures in the following sections can be found at \url{https://github.com/ptoulis/ri-exoplanet-detection}. 
A {\tt Python} implementation can be found at \url{https://github.com/afeinstein20/periodic-ri}.
} Moreover, a similar analysis with a larger range for $\Theta$~(not shown in the figure) reveals multiple peaks ---at periods less than 0.5 million years--- that are even larger than the peaks shown in Figure~\ref{fig:extinction}, and are accepted in the confidence set.
We conclude that there are significant identification issues with this dataset, and so the discovery cannot be confirmed. These issues were also highlighted by~\citet{stigler1987substantial}, albeit 
using a heuristic approach.

\begin{figure}[h!]
    \centering
    \begin{minipage}{0.50\textwidth}
        \centering
        \includegraphics[width=1.08\textwidth]{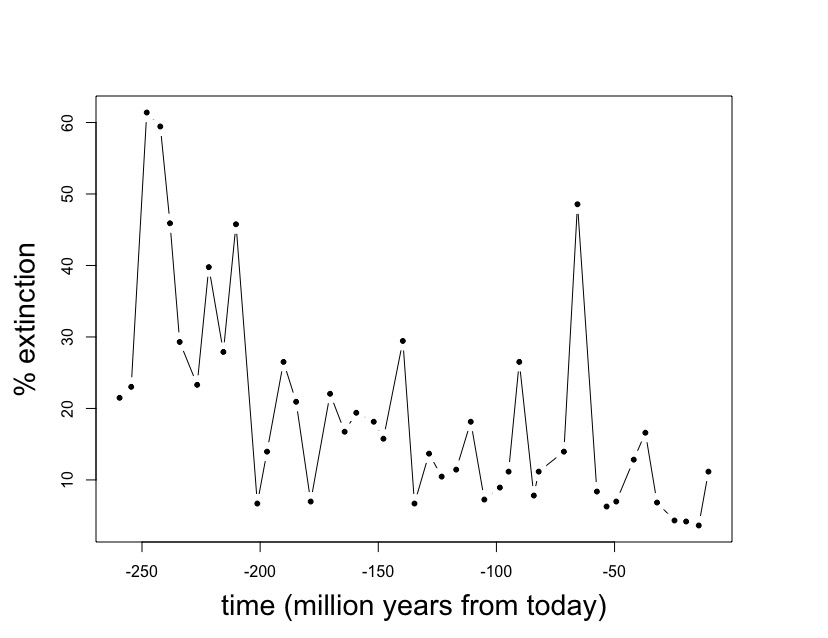} % first figure itself
    \end{minipage}\hfill
    \begin{minipage}{0.5\textwidth}
        \centering
        \includegraphics[width=1.08\textwidth]{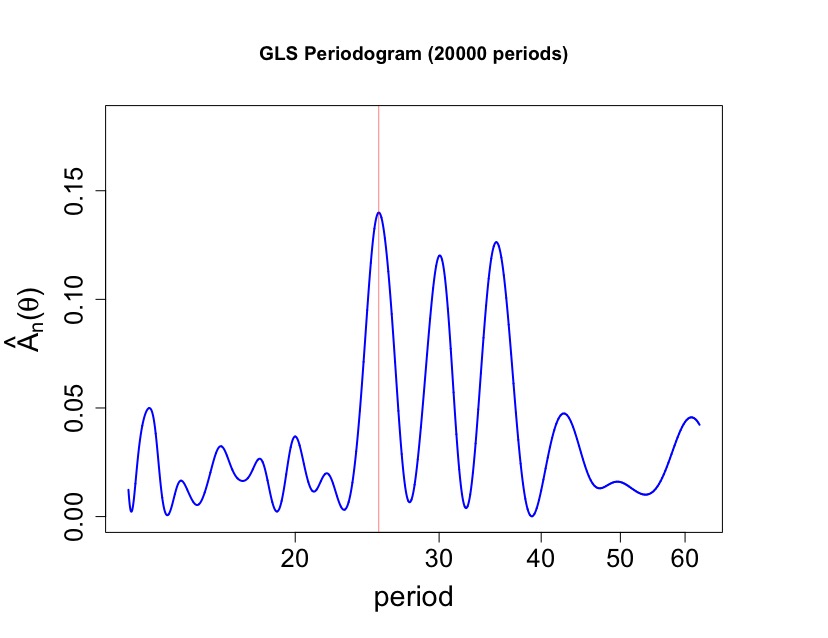} % second figure itself
%        \caption{second figure}
    \end{minipage}
          \caption{\small
          {\em Left:}~Original extinction data of~\citet{raup1986periodic}. 
          The data include 40 observations of extinction events measured as \% of estimated biomass lost during the event.~{\em Right:}~
The periodogram of the extinction data in the range 12.5-62.5 million years, as suggested by~\citet{raup1986periodic}.}
\label{fig:extinction}
\end{figure}

\subsection{Simulated data with realistic observation designs}
Here, we revisit Example~\ref{example1} but with a more realistic design for the observation times.
Specifically, we build upon  the  setup of~\citet[Section 3.2]{hall2006using} and define the data model:
\begin{align}\label{eq:sim1_Y}
y_i = 1-\cos(2\pi t_i/\truetheta) + N(0, \sigma^2),~\quad(i=1,\ldots, n).
\end{align}
Here, the design of observation times relies on a density $f_1$ on the interval $[0, 1]$ that represents a full 24-hour day, with $[0, 0.5]$ denoting day time and $[0.5, 1]$ denoting night time~($t=.75$ is midnight, and so on).  Adapting the setup of~\citet{hall2006using}, we consider three choices for $f_1$:

\begin{enumerate}[(i)]
\item  $f_1(t) = \Ind(.5 \le t \le  1)$;
\item  $f_1(t) = \max\{0, -\pi \sin(2\pi t)\} \Ind(0\le  t \le 1)$;
\item  $f_1(t) = \Ind( | t- 0.75| \le 0.0208)$;
\end{enumerate}
The full observation design is then defined as:
\begin{align}\label{eq:sim1_T}
f(\Tn) \propto \prod_{i=1}^n \left(\sum_{j=0}^{n-1} f_1(t_i - j \truetheta_{\obs})\right).
\end{align}

We set $\truetheta=\sqrt{2}$ as the true signal period with $\sigma^2 =1.5^2$, and $\truetheta_{\obs}=1$ for
the observation period. Thus, the particular form of $f(\Tn)$ in~\eqref{eq:sim1_T} represents a density that has daily periodicity in the interval $[0,n]$, which represents $n$ days of observations.
The definitions of $f_1$ are meant to represent certain recognizable classes of observation designs. 
Specifically, in design (i) the observation times $t_i$ are sampled uniformly during night time.
In design (ii), we define a more realistic scenario where observations happen only at night time, but are more probable during midnight. Finally, in design (iii) the observations are uniformly sampled within one hour window around midnight.

\begin{figure}[t]
\centering
\includegraphics[scale=0.53]{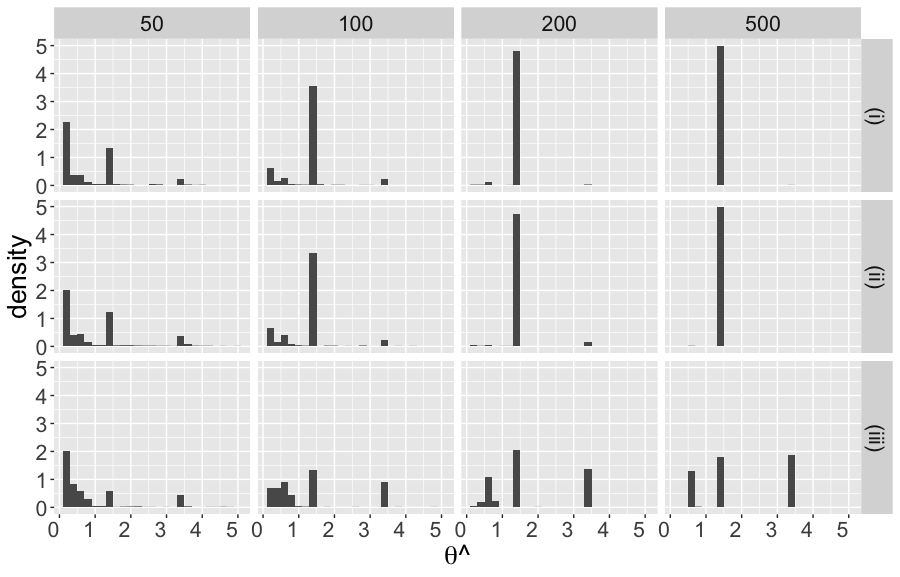}
\caption{\small
Sampling distribution of periodogram peak~(approximated via 1,000 samples) in the model described by
~\eqref{eq:sim1_T} and~\eqref{eq:sim1_Y}.}
\label{fig1}
\end{figure}

Figure~\ref{fig1} shows the sampling distribution of $\hat\theta_n = \arg\max_{\theta\in\Theta} \hat A_n(\theta \mid \Yn, \Tn, \hat\psi_\theta)$, the periodogram peak, where $(\Tn, \Yn)$ are sampled, respectively, according to~\eqref{eq:sim1_T} and~\eqref{eq:sim1_Y}, 
and $\hat A_n$ is the generalized Lombe--Scragle periodogram in~\eqref{eq:real_per}.
As discussed in Section~\ref{sec:intro}, we see that the sampling distribution is far from normal across all designs, 
and is better described as an irregular, discrete distribution on $\Theta$.
The amount of irregularity depends on the observation design.
Under Design (i), for example, $\hat\theta_n$ has converged to a point distribution around $\truetheta$ at $n=500$
data points, but there is significant irregularity even at $n=100$.
Under Design (iii), the situation is much worse since the distribution of $\hat\theta_n$ is irregular 
even at $n=500$ with three distinct modes similar to the mode around $\truetheta$.
This example illustrates one of the key motivations of this paper, which is that standard asymptotic normal theory 
is inappropriate for inference of periodicity in time series due to the 
irregularity of periodic models.

\begin{figure}[h!]
\begin{floatrow}
\centering
\includegraphics[scale=0.315]{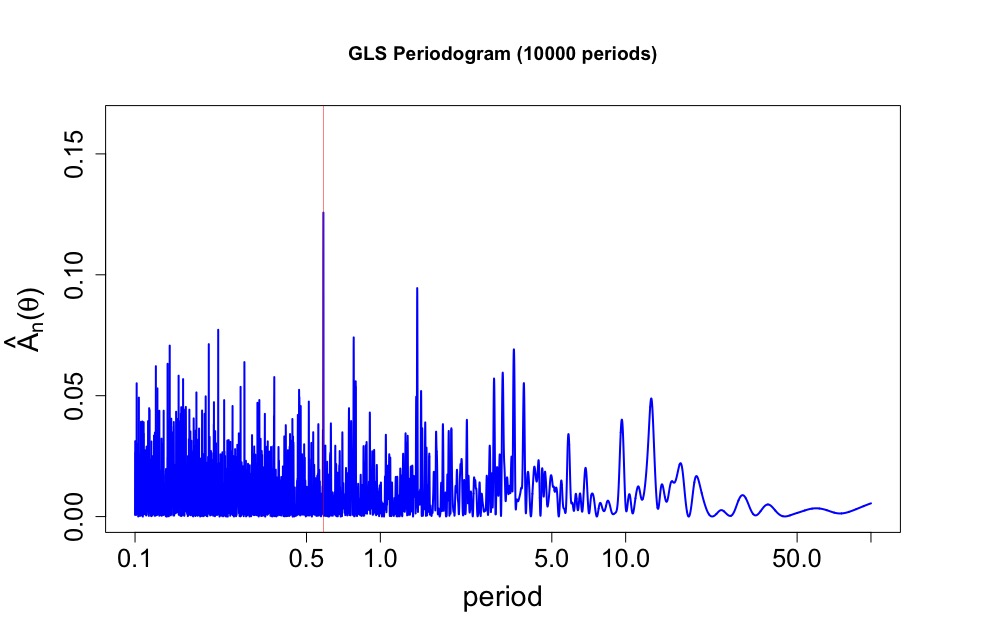}

\capbtabbox{%
  \begin{tabular}{r | r cc}
  $\theta_0$ & $p$-value & $\hat\Theta_{0.95}$ & $\hat\Theta_{0.99}$ \\
   \hline
   0.1216  &  0.0645& yes & yes  \\
   0.1357 &  0.0465 &  no &  yes \\
   0.1385  & 0.0625 &   yes &   yes \\
   0.1998  & 0.0570 &   yes &   yes  \\
   0.2183  & 0.0680 &   yes &   yes  \\
   0.2792  & 0.0440 &  no &   yes  \\
   {\bf 0.5858}  & 0.6460 &   yes &   yes  \\
  0.7787  & 0.0760 &  yes &   yes  \\
 1.4136  & 0.0955 &   yes &   yes  \\
   3.1546  & 0.0675 &  yes &   yes  \\
 9.6474 &  0.0720 & yes & yes \\
 12.7177 & 0.0600 & yes & yes
   \end{tabular}
}{%
}
\end{floatrow} 
%  \vspace{30px}
\caption{\small
{\em Left:} Periodogram for a problematic dataset sampled according to~\eqref{eq:sim1_T} and~\eqref{eq:sim1_Y}, where the peak $\hat\theta_n=0.586$ is not at the true period~$\truetheta=1.414$.
{\em Right:}~Inference of periodicity based on Procedure~1. The table 
shows the $p$-values for the hypothesis $H_0:\truetheta=\theta_0$ for values of $\theta_0$ that correspond 
to the highest peaks of the periodogram shown on the left.
}
\label{fig2}
\end{figure}

Figure~\ref{fig2} shows how to address this problem using  the identification method proposed in this paper. Specifically, following Example~\ref{example1}, we seed a dataset $(\Tn, \Yn)$ such that estimation is problematic 
in the sense that the observed periodogram peak is not at the true value, $\truetheta$~(see Figure~\ref{fig2}, left) ---in this dataset $\hat\theta_n=0.586$. 
The table on the right of Figure~\ref{fig2} shows the $p$-values for testing 
$H_0:\truetheta=\theta_0$ obtained from Procedure 1. For simplicity, we only show results 
for the 10 highest peaks in the observed periodogram.
We see that all these values cannot be rejected at the 1\% level and so they are included in the 
99\% confidence set, $\hat\Theta_{0.99}$~(see column ``$\hat\Theta_{0.99}$" in the table). 
By looking at the range of values included in this confidence set, we conclude that there are 
severe identification issues in the data. This was not immediately evident from the periodogram as the peak in Figure~\ref{fig2} is fairly distinct and far exceeds the 1\% false alarm probability threshold~(not shown above). We also note that, beyond the mode at $\theta_0=0.586$, our proposed method puts the highest confidence at the value $\theta_0=1.4136$~($p$-value = 0.096), which is close to the true signal.

Finally, we check whether the finite-sample coverage of $\hat\Theta_{1-\alpha}$ used in Procedure~1 
matches the asymptotic nominal coverage, as derived in Theorem~\ref{thm3}. 
For this simulation we consider all aforementioned designs ($f_1$), and also multiple sample sizes ($n$), and observation times restricted to a 6-month observation period. For each setting we calculate how often $\hat\Theta_{.95}$ covers the true period, $\truetheta$, over 4,000 replications for each $(f_1, n)$-pair. The results are shown in the table below. 
\renewcommand{\arraystretch}{1.1}
\begin{table}[h!]
\begin{tabular}{c | ccccc}
Design, $f_1$ & \multicolumn{5}{c}{Sample size, $n$} \\
& 25 & 50   & 100  & 250  & 500  \\
\hline
(i)  & 97.97\% &  97.32\%   & 96.05\% & 94.45\% &  95.27\% \\
(ii) & 97.82\% & 97.65\% & 95.42\%  & 94.50\%   & 97.50\%  \\
(iii) & 97.72\% & 98.85\% & 96.95\% & 93.40\% & 93.60\%
\end{tabular}
\label{tab:cover}
\end{table}

We note that coverage gets closer to the nominal value 
for larger sample sizes, which is consistent with Theorem~\ref{thm3}. Moreover, 
we see that better observation designs generally lead to better coverage.

\section{Application: Exoplanet Detection}\label{sec:exoplanet}
In this section, we use our method to analyze radial velocity data from four known star systems. 
%, which some of them have been confirmed independently through other methods (e.g., transits) while the rest remain unconfirmed.
Specifically, we first analyze data on two exoplanets, namely 51 Pegasi b~\citep[``51Pegb"]{mayor1995jupiter} and Gliese 436 b~\citep[``GJ436b"]{butler2004neptune}.
These exoplanets have been confirmed independently by other means, including
direct observations of transit.
%
% We also analyze data from three candidate exoplanets potentially orbiting the stars HD85512~\citep{pepe2011harps}, $\alpha$ Proxima B~\citep{dumusque2012earth}, and Proxima Centauri~\citep{anglada2016terrestrial}, respectively.
We also analyze data from two candidate exoplanets potentially orbiting the stars $\alpha$ Centauri B~\citep{dumusque2012earth}, and Proxima Centauri~\citep{anglada2016terrestrial}, respectively.

\begin{figure}[t!]
\begin{floatrow}
\centering
\includegraphics[scale=0.36]{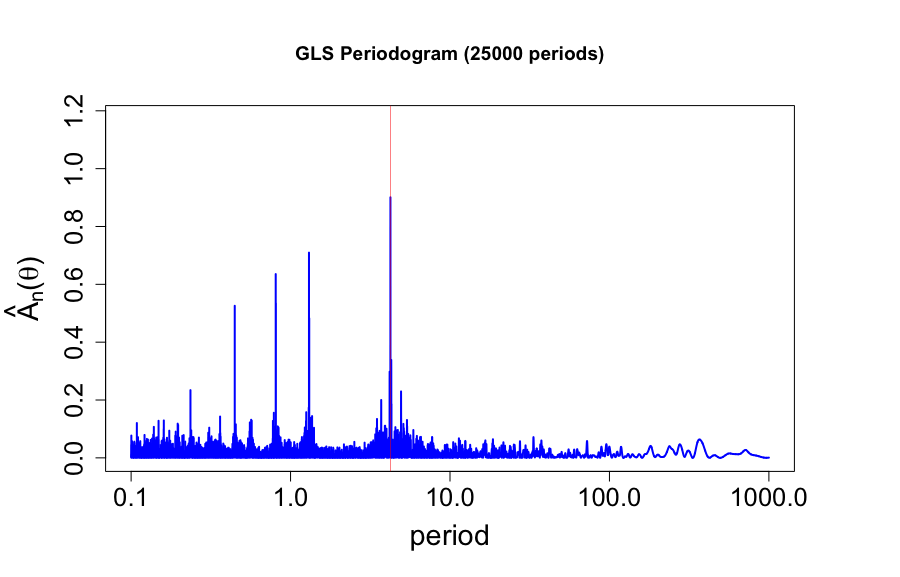}

\capbtabbox{%
  \begin{tabular}{r | r cc}
  $\theta_0$ & $p$-value & $\hat\Theta_{0.95}$ & $\hat\Theta_{0.99}$ \\
   \hline
0.3085 & 0.0000 & no & no \\ 
  0.5662 & 0.0000 & no & no \\ 
  0.8069 & 0.0000 & no & no \\ 
  0.8089 & 0.0000 & no & no \\ 
  0.8295 & 0.0000 & no & no \\ 
  1.3047 & 0.0000 & no & no \\ 
  1.3095 & 0.0000 & no & no \\ 
  3.7033 & 0.0000 & no & no \\ 
  4.1807 & 0.0000 & no & no \\ 
 {\bf  4.2311} & 1.0000 & yes & yes \\ 
  4.2821 & 0.0000 & no & no \\ 
  4.9331 & 0.0000 & no & no \\ 
   \end{tabular}
}{%
}
\end{floatrow} 
%  \vspace{30px}
\caption{\small
{\em Left:}~Periodogram of radial velocity on exoplanet ``51Pegb". Here, $\Theta = \{0.1, \ldots, 1000\}$ is split uniformly in the log-space so that 
$|\Theta| = 25,000$. 
{\em Right:}~Inference of periodicity of 51Pegb~based on Procedure~1.
The table shows the $p$-values for the hypothesis $H_0:\truetheta=\theta_0$ for values of $\theta_0$ that correspond to high peaks of the periodogram shown on the left.
We see that there are no identification issues, as the 4.23-day signal is the only one accepted 
in the confidence sets.
\vspace{25px}
}
\label{fig3}
\end{figure}

We start our analysis with the confirmed exoplanets in our sample. The periodogram of the stellar radial velocity data for 51Pegb is depicted in Figure~\ref{fig3}. 
This shows a clear peak around $\hat\theta_n=4.23$ days with additional 
distinct peaks around 1-day and 0.5-day periods.
Our method correctly shows no identifiability issues here.
All periods corresponding to the highest peaks are strongly rejected ($p$-values numerically zero) except for 
$\theta_0=4.23$ (observed peak).
Next, we analyze data on GJ436b. Compared to 51Pegb, the periodogram for GJ436b shows  a peak around $\hat\theta_n=2.6441$, but this peak is less distinct than other peaks. Our method again correctly shows  that the true signal can be easily identified.
As in 51Pegb,  all periods corresponding to the highest peaks are strongly rejected ($p$-value = 0) except for the observed peak at $\theta_0=2.64$.

\begin{figure}[h!]
\begin{floatrow}
\centering
\includegraphics[scale=0.36]{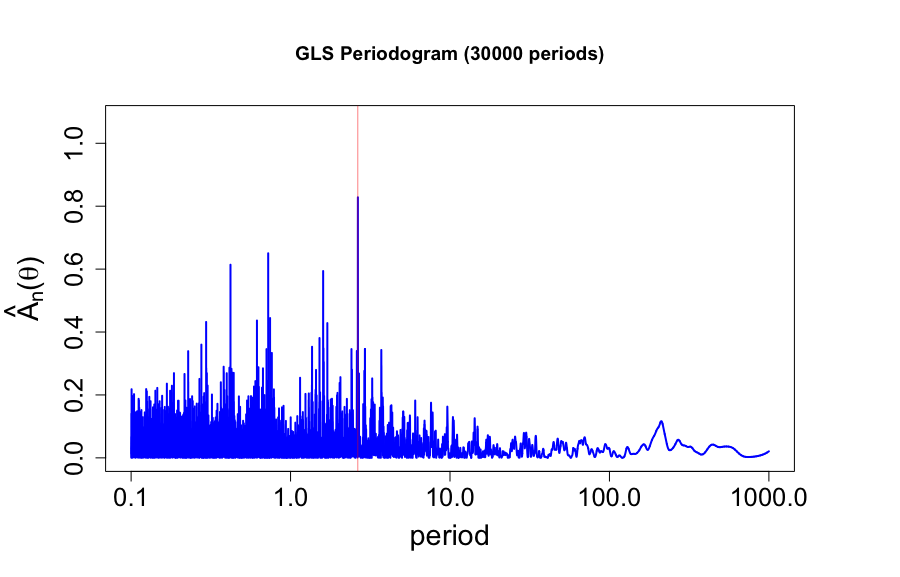}

\capbtabbox{%
  \begin{tabular}{r | r cc}
  $\theta_0$ & $p$-value & $\hat\Theta_{0.95}$ & $\hat\Theta_{0.99}$ \\
   \hline
0.4200 & 0.0000 & no & no \\ 
  0.6155 & 0.0000 & no & no \\ 
  0.7067 & 0.0000 & no & no \\ 
  0.7438 & 0.0000 & no & no \\ 
  1.3641 & 0.0000 & no & no \\ 
  1.5187 & 0.0000 & no & no \\ 
  1.6013 & 0.0000 & no & no \\ 
  1.6086 & 0.0000 & no & no \\ 
  1.7008 & 0.0000 & no & no \\ 
  2.4103 & 0.0000 & no & no \\ 
 {\bf  2.6441} & 1.0000 & yes & yes \\ 
  3.7092 & 0.0000 & no & no \\ 
   \end{tabular}
}{%
}
\end{floatrow} 
%  \vspace{30px}
\caption{\small
{\em Left:}~Periodogram of radial velocity on exoplanet ``GJ436b". Here, $\Theta = \{0.1, \ldots, 1000\}$ is split uniformly in the log-space so that 
$|\Theta| = 30,000$. 
{\em Right:}~Inference of periodicity of GJ436b~based on Procedure~1.
The table shows the $p$-values for the hypothesis $H_0:\truetheta=\theta_0$ for values of $\theta_0$ that correspond to high peaks of the periodogram shown on the left.
We see that there are no identification issues, as the 2.64-day signal is the only one accepted 
in the confidence sets.
\vspace{20px}
}
\label{fig4}
\end{figure}

\begin{figure}[t!]
\begin{floatrow}
\centering
\includegraphics[scale=0.35]{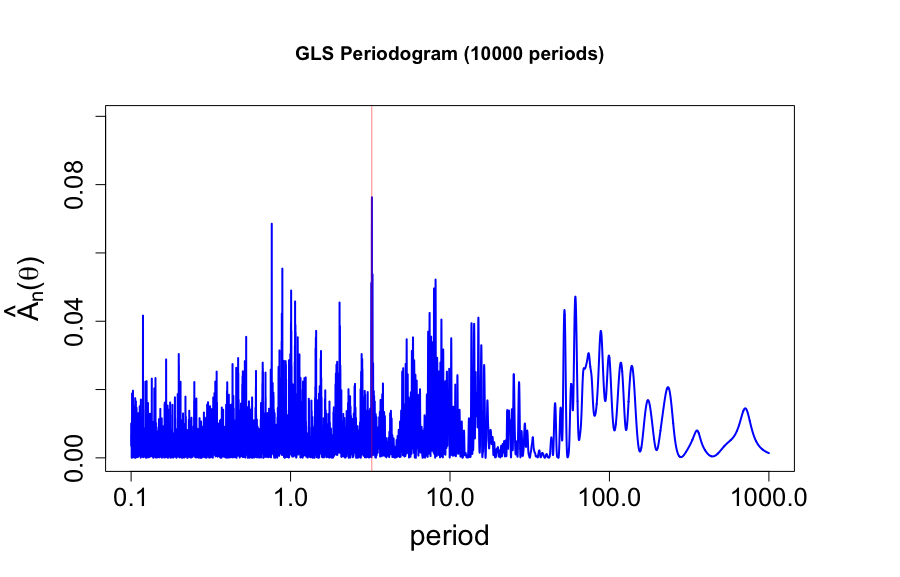}

\capbtabbox{%
  \begin{tabular}{r | r cc}
  $\theta_0$ & $p$-value & $\hat\Theta_{0.95}$ & $\hat\Theta_{0.99}$ \\
   \hline
0.7622 & 0.0705 & yes & yes \\ 
  0.8882 & 0.0271 & no & yes \\ 
  1.0086 & 0.0174 & no & yes \\ 
  1.0678 & 0.0079 & no & no \\ 
  2.0292 & 0.0122 & no & yes \\ 
  3.2074 & 0.0163 & no & yes \\ 
  {\bf 3.2371} & 1.0000 & yes & yes \\ 
  3.2670 & 0.0178 & no & yes \\ 
  7.9394 & 0.0116 & no & yes \\ 
  8.1169 & 0.0175 & no & yes \\ 
  52.2242 & 0.0121 & no & yes \\ 
  61.1334 & 0.0226 & no & yes \\ 
   \end{tabular}
}{%
}
\end{floatrow} 
%  \vspace{30px}
\caption{\small
{\em Left:}~Periodogram of radial velocity on candidate exoplanet orbiting $\alpha$ Centauri B~\citep{dumusque2012earth}.  Here, $\Theta = \{0.1, \ldots, 1000\}$ is split uniformly in the log-space, so that 
$|\Theta| = 10,000$. 
{\em Right:}~Inference of periodicity of the exoplanet~based on Procedure~1.
The table shows the $p$-values for the hypothesis $H_0:\truetheta=\theta_0$ for values of $\theta_0$ that correspond to high peaks of the periodogram shown on the left.
We see that there are severe identification issues as several signals other than the periodogram peak are accepted in the confidence sets.\vspace{20px}}
\label{fig5}
\end{figure}

Our results are mixed for the other, yet unconfirmed exoplanets.
For the candidate exoplanet orbiting $\alpha$ Centauri B~\citep{dumusque2012earth}, in particular,
our analysis ---summarized in Figure~\ref{fig5}--- shows severe identification issues.
The periodogram shows a peak around $\hat\theta_n=3.23$ days but it is not as distinct 
from the other peaks. Moreover, there is significant variation in the periodogram. 
Indeed, our method reveals that many periods corresponding to the highest peaks cannot be rejected at the 1\% level, and there is even a period ($\theta_0=0.7622$) that cannot be rejected  at the 5\% level.
These results indicate that additional statistical evidence is necessary to claim that the particular exoplanet discovery is plausible.
As in the case of the extinction data, we note that  earlier work 
has raised concerns about this discovery as well~\citep[e.g.]{rajpaul2015ghost}, 
albeit using only  heuristic methods similar to those suggested by~\citet{stigler1987substantial}.

\begin{figure}[h!]
\begin{floatrow}
\centering
\includegraphics[scale=0.35]{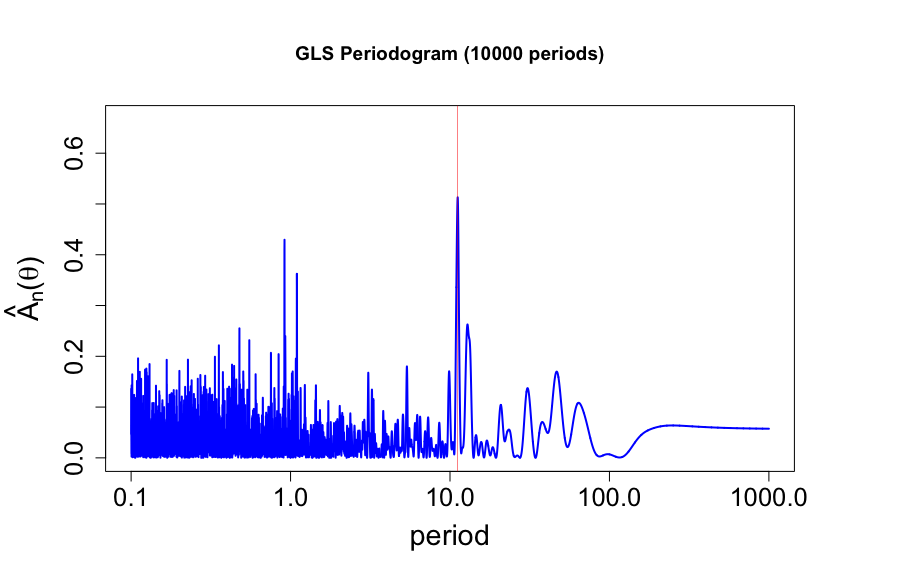}

\capbtabbox{%
  \begin{tabular}{r | r cc}
  $\theta_0$ & $p$-value & $\hat\Theta_{0.95}$ & $\hat\Theta_{0.99}$ \\
   \hline
0.1106 & 0.0007 & no & no \\ 
  0.3355 & 0.0022 & no & no \\ 
  0.3552 & 0.0055 & no & no \\ 
  0.4778 & 0.0025 & no & no \\ 
  0.5512 & 0.0047 & no & no \\ 
  0.7532 & 0.0052 & no & no \\ 
  0.8412 & 0.0059 & no & no \\ 
  0.9164 & 0.0173 & no & yes \\ 
  0.9266 & 0.0005 & no & no \\ 
  1.0957 & 0.0080 & no & no \\ 
  {\bf 11.1739} & 1.0000 & yes & yes \\ 
  12.8769 & 0.0006 & no & no \\ 
   \end{tabular}
}{%
}
\end{floatrow} 
%  \vspace{30px}
\caption{
\small
{\em Left:}~Periodogram of radial velocity on candidate exoplanet Proxima Centauri b~\citep{anglada2016terrestrial}.  Here, $\Theta = \{0.1, \ldots, 1000\}$ split regularly in the log-space, so that 
$|\Theta| = 10,000$. 
{\em Right:}~Inference of periodicity of the exoplanet~based on Procedure~1.
The table shows the $p$-values for the hypothesis $H_0:\truetheta=\theta_0$ for values of $\theta_0$ that correspond to high peaks of the periodogram shown on the left.
We see that there are no severe identification issues, and the detection appears to be robust.
There is only only signal (at 0.9164 days) that cannot be rejected at the 1\% level, and thus causes concern.
An improved observation design could decisively rule out this nuisance signal; see Section~\ref{sec:design} for 
more details.\vspace{20px}
}
\label{fig6}
\end{figure}

Finally, we discuss the data for a recent discovery of an exoplanet ostensibly orbiting 
Proxima Centauri~\citep{anglada2016terrestrial}.
We use a recent release of 63 observations of Proxima Centauri from the ESPRESSO spectograph~\citep{mascareno2020revisiting} that has increased precision.
The analysis and results are shown in Figure~\ref{fig6}. 
The periodogram shows  a peak around $\hat\theta_n=11.17$ days but also 
a substantial peak around the 1-day period signal, possibly  associated with the observation design.  Indeed, our method shows that out of all highest peaks in the periodogram, only 
$\theta_0=11.1739$ and $\theta_0=0.9164$ cannot be rejected at the 1\% level. 
% Another peak, namely $\theta_0=1.0957$, has a $p$-value of 0.0080, which while it is smaller than 1\% it still may be considered high for the purposes of scientific discovery.
Overall, while the finding of \citet{anglada2016terrestrial} seems to be robust, there is still a small chance of observing the observed periodogram pattern under the particular observation times, at least with respect to those aspects of the periodogram that are captured by our test statistic.
More statistical evidence is therefore needed to statistically validate this discovery, 
possibly through the use of a slightly more randomized observation design. 
We discuss such designs in the  following section.
%
%
%\begin{figure}[t]
%\centering
%\includegraphics[scale=0.52]{fig/fig1_sampling2}
%\caption{sampling distirbutio over 1,000 samples \todo{complete}}
%\label{fig1}
%\end{figure}
%

\section{Discussion and Extensions}\label{sec:extensions}

\subsection{Observation design}\label{sec:design}

The importance of observation times in identifying a periodic signal is well understood~\citep{feigelson2012modern, vanderplas2018understanding, ivezic2014statistics}.
It may therefore come as a surprise that  there is little systematic work on optimizing observation designs.
One exception is perhaps~\citep{loredo2003bayesian}, which discusses a Bayesian approach to 
sequential experimental design of observations---see also~\citep{wynn1984jack,kiefer1975optimal, mitchell2000algorithm,smucker2018optimal} for a statistical background.
However, this approach also relies on regularity conditions for the underlying data model, and so it does not account for
the identification issues  discussed in Section~\ref{sec:intro}.

One way to address these issues is to use the identification method developed in this paper.  The idea is simply to synthesize data under alternative designs, and then pick the design 
that yields the most precise confidence set, $\hat\Theta_{1-\alpha}$. 
Conceptually, the main challenge with this approach is how to define an appropriate space for these alternative designs in a way that is both practical and computationally efficient.
Given a choice for the design space, we can then simulate synthetic data 
and choose from those designs that yield  narrow confidence sets.
% One straightforward way to quantify the precision of a confidence set is through the range of values contained in the set.

In practice, we usually face a scenario where we suspect the peak $\hat\theta_n$ in the periodogram to be the true signal, but there is also another ``nuisance peak", say $\thetanui$. 
Typically, this nuisance peak appears near the 1-day signal as a by-product of a daily observation design.
For example, in the ESPRESSO data discussed in Section~\ref{sec:exoplanet} and Figure~\ref{fig6}, the problem is that 
even though there is very strong evidence for a signal at $\hat\theta_n=11.17$ days, 
there is also a signal at $\thetanui =0.9164$ that cannot be rejected at the 1\% level~($p$-value = 0.017). The observed peak could statistically be an artifact of the daily observation design. 
The key question is thus: 
\begin{quote}
``What kind of observation designs should we use to accurately 
identify the signal at $\hat\theta_n$, {\em assuming} this signal to be the true one?"
\end{quote}

To address this problem, we propose the following high-level procedure.

\begin{enumerate}
\item Select a space $\mathcal{D}$ for observation designs and a tolerance threshold $0 < \epsilon < |\hat\theta_n - \thetanui|$.
\item {\bf For} each design $D$ in the design space, {\bf do:}
\begin{enumerate}[(a)]
\item Assume $\truetheta= \hat\theta_n$ as the true signal, and then calculate $\hat\psi_{\truetheta}$ as defined in~\eqref{eq:real_per}.
\item {\bf For} $r=1, \ldots, R$, {\bf do:}
\begin{enumerate}[(i)]
\item Generate synthetic data $(\TTnn, \TYnn)$ according to the design and $(\truetheta, \hat\psi_{\truetheta})$ from step 2(a), where  we may 
introduce additional (synthetic) observations, such that $n' \ge n$.
\item  Calculate $\hat\Theta_{1-\alpha}$ based on synthetic data $(\TTnn, \TYnn)$.
\item Calculate how many periods in $\hat\Theta_{1-\alpha}$ are $\epsilon$-far from $\truetheta$:
$$
I_{D}^r = \mid \{ \theta\in\hat\Theta_{1-\alpha} : |\theta - \truetheta| > \epsilon \} \mid.
$$
%  and calculate synthetic radial velocity measurements $\TYn = [ \yp(\tilde t_i; \hat\psi, \truetheta) +\eps(\tilde t_i)]$,  where the errors can be sampled i.i.d. $N(0, \sigma_i^2)$.
\end{enumerate}
\end{enumerate}
\item Select the ``simplest" design possible that also yields sharp identification:
\begin{align}\label{eq:best_design}
\hat{D} = \arg\min_{D\in\mathcal{D}} 
\left\{ |D | : \sum_{r=1}^R I_{D}^r = 0\right\},
\end{align}
where $|D| \in\mathbb{R}$ denotes the complexity of the design~(explained below).
\end{enumerate}

In words, the design $\hat D$ in~\eqref{eq:best_design} is the simplest possible observation design 
for which the confidence sets include only values in the $\epsilon$-neighborhood of $\truetheta$.
In this case, we can accurately identify the underlying true signal. To implement the optimal design procedure outlined above, we only have to define the design space $\mathcal{D}$ along with an appropriate notion of design complexity, and also discuss choices for threshold $\epsilon$. Here, for simplicity, we consider two choices for the design space:
\begin{enumerate}[(A)]
\item {\bf Fixed observations, $n$}.  
Each design in this space is parameterized by $\delta>0$ such that 
$\TTn = \Tn + \delta U^{(n)}$, where $U^{(n)}$ is a vector of $n$ independent $\Unif([-1, 1])$, 
i.e., random variables uniformly distributed on $[-1,1]$.
Thus, the alternative observation time $\tilde t_i$ is uniformly distributed $\pm \delta$ days around 
the actual observation time $t_i$. 
Given $\TTn$, we may generate synthetic radial velocity data, $\TYn$, through
$y(\tilde t_i) = \yp(\tilde t_i; \truetheta, \hat\psi_{\truetheta}) + \eps_i$, 
where $\yp$ is the periodic component defined in~\eqref{A2}, and $\eps_i \sim N(0, \sigma_i^2)$, i.i.d.
The complexity of a design $D$ with parameter $\delta$ can simply be defined as
$| D| = \delta$.

\item {\bf Additional observations, $n' - n$.}
Each design in this space is parameterized by $n'>0$ such that 
$\TTnn = (\Tn, \tilde T^{(n'-n)})$, where we sample $(n'-n)$ additional observation times in a hypothetical span of 3 months as follows:
$\tilde t_j = \max \{\Tn\} + \Unif(\{1,\ldots, 90\}) + 0.01 \cdot \Unif([-1,1])$, for $j=1, \ldots, (n'-n)$.
Given $\TTnn$, we can then generate synthetic radial velocity data $\TYnn$ exactly as in Design (A).
The complexity of a design $ D$ can simply be defined as
$| D| = n'-n$.
\end{enumerate}

Intuitively, Design (A) adds some level of randomization (controlled by $\delta$) on top of the actual observation times $\Tn$. Roughly speaking, this determines the amount of randomization necessary for accurate identification in future observations. When there are identification issues, a large $\delta$ will generally be necessary for identification, and, conversely, when there are no identification issues a small $\delta$ (or zero) will be needed.
Design (B) adds new observation time points at a pre-specified level of noise. 
This level is set to $\pm 0.01$ on the day-scale which amounts to $2\times 0.01\times 24 \times 60 \approx$ 30 mins of randomized observation window around $n'-n$ additional daily observations.
Similar to Design (A), when there are identification issues, a large number of additional observations, $n'-n$, will be necessary for identification; and, conversely, when there are no identification issues, a small number of 
additional observations will be needed.

Regarding the hyper-parameters, we note that $R$ controls the number of synthetic datasets 
generated per design, and thus large $R$ leads to more accurate calculation of the optimal design in~\eqref{eq:best_design}. Practically, since we want to estimate small, near-zero proportion values, $R$ could be small, e.g., $R=1,000$ or  even $R=100$.
Moreover, a reasonable value for $\epsilon$ is in the range $[0.1, 0.5]$. Increasing that threshold to larger values could conflate real small-periodicity signals with the common 1-day period signal that is associated with the daily observation designs.
 
To illustrate these ideas, we apply the above optimal design procedure to the exoplanet detection problems of Section~\ref{sec:exoplanet}. The results are shown in Table~\ref{tab:design}. We see 
 that for 51 Pegasi b and Gliese 436 b, there is no need for improving the observation times. 
 Indeed, we need neither additional randomness in observation times ($\delta=0$), 
 nor any additional observations ($n'-n=0)$. 
 In contrast, for the candidate exoplanet around $\alpha$ Centauri B~\citep{dumusque2012earth}
 we need a substantial improvement to the observation design. 
Specifically, in order to identify precisely the observed signal at $\hat\theta_n=3.23$ at the 1\% level~(assuming it's true),  we need a large additional variation of $\pm 0.18$ days around the actual observation times 
(i.e., $\pm 4.32$ hrs./observation). Alternatively, we need 137 additional observations with a random variation of $\pm 15$ mins./observation, as specified by Design (B), almost equal to the size of the actual dataset.
Finally, the situation is much better for Proxima Centauri~\citep{anglada2016terrestrial}. 
Specifically, in order to identify the observed signal at $\hat\theta_n=11.17$ at the 1\% level ---and so eliminate the only nuisance signal at $\thetanui=0.9164$---  we need an additional variation of $\pm 0.06$ days~(i.e., $\pm 1.44$ hrs./observation) around the actual observation times. Alternatively, we only need 17 additional observations with a random variation of 
$\pm 15$ mins./observation, as specified by Design (B). 

In summary,  the results in Table~\ref{tab:design} illustrate the benefits of the design procedure outlined above in guiding the selection of observation times for future exoplanet detections. In practice, we can envision a two-stage approach where at the first stage the scientist analyzes the stellar signal through the periodogram and identifies the most plausible peak along with other nuisance peaks. In the second stage, the scientist performs an analysis as described here in order to quantify how much variation or how many more samples are needed for identification of the plausible signal at a desired significance level, and proceeds to make these observations.

Of course, one limitation of our discussion here is that we have only considered two simple classes of observations designs. In practice, we need to also take into account the particular logistics 
of Earth-based observations to determine more realistic observation designs. 
The optimal design procedure in~\eqref{eq:best_design} could then be used in conjunction with such more realistic designs. Overall, this is an intriguing statistical problem, which we suggest for future work.

\renewcommand{\arraystretch}{1.3}
\begin{table}[t!]
\small
\begin{tabular}{lccc}
& \multicolumn{2}{c}{Design (A)} & Design (B) \\
\hline
\multicolumn{1}{l|}{Candidate Exoplanet (or Star)}                                                  & \begin{tabular}[c]{@{}c@{}} \small{randomness needed} \\ \small{for identification} (best $\delta$)
\end{tabular} & \multicolumn{1}{c|}{ \small{$\pm$ hrs.}} & \begin{tabular}[c]{@{}c@{}} \small{\#additional obs.~needed} \\
\small{for identification~(best $n'-n$)}
\end{tabular} \\ \hline
\multicolumn{1}{l|}{51 Pegasi b~\footnotesize{\citep{mayor1995jupiter}  }   }      & 0                                                                         & \multicolumn{1}{c|}{0}                   & 0                                                                                  \\
\multicolumn{1}{l|}{Gliese 436 b~\footnotesize{\citep{butler2004neptune}}}           & 0                                                                         & \multicolumn{1}{c|}{0}                   & 0                                                                                  \\
\multicolumn{1}{l|}{$\alpha$ Centauri B~\footnotesize{\citep{dumusque2012earth}}}     & 0.18                                                                      & \multicolumn{1}{c|}{4.32}                & 137                                                                            \\
\multicolumn{1}{l|}{Proxima Centauri\footnotesize{~\citep{anglada2016terrestrial}}} & 0.06                                                                      & \multicolumn{1}{c|}{1.44}                & 17                                                                             
\end{tabular}
\caption{
\small Observation designs (A) and (B) to achieve identification 
in the exoplanet applications of Section~\ref{sec:exoplanet}.
Design (A) introduces additional randomness in the observation times, while design (B) introduces additional observations. 
%They both quantify how to improve the observation times to achieve identification.
%
We see that 51Pegb and GJ436b require no improvement in the observation times as the signal is well-identified.
To identify the candidate exoplanet around $\alpha$ Centauri B, however, 
  we need an additional variation of $\pm 0.18$ days around the actual observation times 
(i.e., $\pm 4.32$ hrs./observation). 
Alternatively, we need 137 additional observations with a random variation of $\pm 15$ mins./observation.
For the candidate exoplanet around Proxima Centauri~\citep{anglada2016terrestrial} we need an additional variation of $\pm 0.06$ days~(i.e., $\pm 1.44$ hrs./observation) on the actual observation times. Alternatively, we only need 17 additional observations with a random variation of 
$\pm 15$ mins./observation.%as specified by Design (B).  
}
\label{tab:design}
\end{table}

\subsection{Non-parametric approach}\label{sec:np}
The method developed in Section~\ref{sec:method} is parametric and employs the
harmonic model for the periodic component~\eqref{A2}. While this can be extended to include other periodic models (e.g., Keplerian model) it is interesting to consider non-parametric approaches as well. While nonparametric approaches have been considered in the literature before~\citep{feigelson2012modern, ivezic2014statistics, hall2000nonparametric, hall2006using, hall2008nonparametric}, their asymptotics require strong assumptions on the observation design (typically i.i.d observation times), which is unrealistic. As such, nonparametric methods face the  the same identification problems tied to irregular observation times as parametric methods.
For example, the identification of a 26-million year biological extinction cycle 
in Example~\ref{example2} used a nonparametric randomization test. However, as discused in Section~\ref{sec:extinction}, the period cannot be identified from this dataset.

Here, we discuss a nonparametric variant of our set identification method. As before, we assume 
that the signal is decomposed as $y_i = \yp(t_i) + \eps(t_i)$, but make no parametric assumption about $\yp$. Before we define and test our main hypothesis, we need some notation.
For any two time points $t, t'$ and period $\theta\in\Theta$, let $t \equiv t'(\mod\theta)$ denote that 
$t-t' = k\theta$ for some integer $k$. Also, let $\mathsf{S}_n$ denote the symmetric group of permutations of $n$-element vectors.
For some fixed $\theta$, define:
$$
\Pi(\Tn; \theta) = \{\pi \in\mathsf{S}_n : t_{\pi(i)} \equiv t_i(\mod\theta),~i=1, \ldots, n\}.
$$
In words, $\Pi(\Tn; \theta)$ is the set of permutations of $\{1, \ldots, n\}$ such that  
any time $t_i$ is mapped only to an observation time that is equivalent to $t_i$ modulo $\theta$.
It is straightforward to see that $\Pi(\Tn; \theta)$ is a subgroup of $\mathsf{S}_n$ since the $\mod$operator defines an equivalence relationship. Let us assume that 
for any $\Tn$ and period $\theta\in\Theta$, the errors satisfy the invariance Assumption~\eqref{A1} with $\GG = \Pi(\Tn; \theta)$. This assumption is generally mild. It requires that the data are exchangeable modulo $\theta$, and actually allows certain correlated structures for the errors $\eps(t)$.

We wish to test the following nonparametric null hypothesis of periodicity $\theta_0$:
\begin{align}
\Hnp:~\yp(t') = \yp(t),~\text{for all}~t',t~\text{such that}~t'\equiv t(\mod\theta_0).\nn
\end{align}
To test $\Hnp$ we can adapt Procedure 1 as follows.
%% PROCEDURE 2
\MakeProcedure{2}{
\begin{enumerate}
% \item Calculate the observed value, $\sobs$, of the test statistic.
\item {\bf For} all $r = 1, \ldots, R$ {\bf do:}
\begin{enumerate}[(i)]
\item Sample $\pi\sim\Unif\big(\Pi(\Tn; \theta_0)\big)$.
\item Generate synthetic outcome data $\Ynr = \pi \cdot \Yn$ obtained by permuting the data $\Yn$ according to $\pi$ while the observation times, $\Tn$, are fixed.
\end{enumerate}
\item Using the samples  from 2(ii), calculate the $p$-value, say $\pval(\theta_0)$, as in~\eqref{eq:pval_theta}, and 
reject if the $p$-value is less than $\alpha$.
\end{enumerate}
}

In words, to test $\Hnp$, Procedure~2, first, picks a permutation of the observation times in $\Tn$ such that 
the permuted times are equivalent to the corresponding original periods modulo $\theta_0$.
Then, it uses this permutation to generate a new outcome vector $\Ynr$. In the last step, the procedure recalculates the test statistic using data $(\Tn, \Ynr)$, which can then be used 
in the calculation of the $p$-value.

\begin{theorem}\label{thm:np}
\ThmFour
\end{theorem}

According to Theorem~\ref{thm:np}, the test of Procedure~2 is exact without requiring a parametric model for the periodic component, $\yp$. This is possible because under $\Hnp$ and the invariance assumption on the errors, any two 
observations $y(t), y(t')$ are exchangeable when $t, t'$ are equivalent modulo $\theta_0$.

Despite the strength of this result, one practical difficulty with Procedure 2 is that it requires 
us to find several pairs of distinct time points in $\Tn$ that are periodically equivalent modulo $\theta$, for all $\theta\in\Theta$. This may be unrealistic in practical applications, including exoplanet detection, because the sample sizes can be small. Technically, the problem is that $\Pi(\Tn; \theta)$ may be a singleton group (including only the identity permutation) for many values of $\theta$, so that Procedure 2 leads to an exact by ``empty" test, and thus to confidence sets that are too conservative. 
To address this problem, it may be tempting to expand the definition of $\Pi(\Tn; \theta)$
to include permutations between $t,t'$ that are $\epsilon$-away from being periodically equivalent modulo $\theta$. Under this approach, however, $\Pi(\Tn; \theta)$ is not a group anymore.
%, and so the test is not valid but may be approximately so. 
%
An alternative approach would be to use the nonparametric estimators of $\truetheta$ 
developed by~\citet{hall2000nonparametric, hall2006using, hall2008nonparametric} together with a variation of Procedure~1 or Procedure~2. Both these procedures do not require regularity conditions on the observation times but only a consistent estimator for the periodic component, $\yp$. We leave these directions for future work.

\section{Conclusion}
We developed a method of set identification for hidden periodicity in unequally spaced time series. 
Our approach is more appropriate than standard methods of statistical inference because common estimators, such as the periodogram peak, are not well-behaved and may even be inconsistent without 
strong (and practically unrealistic) assumptions on the observation design.
%
% Empirically, we illustrate our contributions in two ways. First, we show through a numerical example that the sampling distribution of periodogram peaks can be highly irregular and deviate significantly from normality, which invalidates standard methods of inference, including bootstrap- and Bayesian-based procedures.
We illustrated empirically our method in examples from  exoplanet detection using radial velocity data.
In the confirmed exoplanets of our sample, our method can identify the hidden periodicity precisely. 
We also find that the statistical evidence in a recent candidate around  $\alpha$ Centauri B  is particularly weak, as several nuisance signals cannot be rejected even at the 5\% level. In another popular discovery around Proxima Centauri, the evidence appears to be robust but a nuisance signal due to the daily observational cycle can still not be rejected at the 1\% level.
Our method suggests ways to improve the observation designs, either by introducing 
some randomness in the observation times, or by just making more observations.
These improved designs could resolve the aforementioned identification issues, and also help in 
efficient scheduling of measurements for future discoveries. %The approach is to quantify either how much additional randomness is needed in the observation times or how many more observations are needed in order to identify a candidate signal. 
% Following this approach, we propose observation designs that could address the identification issues in the aforementioned examples of exoplanet detection.

\section{Acknowledgements}
We would like to thank Yang Chen, Eric Feigelson, Vinay Kashyap, Xiao-Li Meng, Aneta Siemiginowska, Katja Smetanina, Michael Vogt, and the participants at the Statistics Department Seminar at UC Berkeley and the ``Topics in Astrostatistics" Seminar at Harvard University for useful suggestions and feedback.

\small
\bibliography{astro}
\bibliographystyle{apalike}

\newpage
\appendix 

\setcounter{theorem}{0}

\section*{A. Proof of Theorem~\ref{thm1}}

\begin{align}
\label{A1}
\gg\cdot \epsn & \eqd \epsn~\mid \Tn~\quad(\gg\in\GG).\tag{A1} \\
\label{A2}
\yp(t; \truetheta, \truepsi) & = \psi^\ast_{1} + \psi^\ast_{2} \cos(2\pi t/\truetheta) + \psi^\ast_{3} 
\sin(2\pi t/\truetheta).
\tag{A2}
\end{align}

The definition of the $p$-value for $\Hfull$ is as follows:
\begin{align}
\pval(\theta_0, \psi_0) = E\{s_n(\Yni, \Tn) \ge \sobs \mid \Tn\},\nonumber
\end{align}

\begin{theorem}\label{thm1}
\ThmOne
\end{theorem}
\begin{proof}
Recall that $\Yni = \Ypn + G^{(i)} \epsn$ where $G^{(i)}\sim \Unif(\GG)$ and $\epsn = \Yn - \Ypn$.
Under $\Hfull$ and Assumptions~\eqref{A1}-\eqref{A2} we obtain
$\Yni \eqd \Yn \mid \Tn$. It follows that
$$
s_n(\Yni, \Tn) \eqd \sobs \mid \Tn
$$
and so the $p$-value in~\eqref{eq:pval_full} is finite-sample valid, that is,
$\EX\big[ \Ind\{\pval(\theta_0, \psi_0) \le \alpha\} \mid \Hfull, \Tn\big] = \alpha$.
It follows that the $p$-value is also valid {\em unconditionally}:
\begin{align}
\EX\big[ \Ind\{\pval(\theta_0, \psi_0) \le \alpha\} \mid \Hfull\big]
= 
\EX\big[~\EX\big[ \Ind\{\pval(\theta_0, \psi_0) \le \alpha\} \mid \Hfull, \Tn\big] \mid \Tn\big]
= \alpha.\nn
\end{align}
\end{proof}

\section*{B. Proof of Theorem~\ref{thm2}}
\begin{align}
& H_0: \truetheta  =\theta_0. 
\nonumber\\
 & \Theta_{1-\alpha} = \{\theta \in \Theta :   \max_{\psi\in\Psi}  \pval(\theta, \psi) > \alpha \}.\nonumber
\end{align}

\begin{theorem}\label{thm2}
\ThmTwo
\end{theorem}
\begin{proof}
Suppose that $H_0$ in~\eqref{eq:H02} holds., Then,
\begin{align}
\pr(\theta_0\notin\Theta_{1-\alpha} \mid H_0)  = \EX[\Ind\{\max_\psi \pval(\truetheta, \psi) \le \alpha\} ]
\le \EX[\Ind\{\pval(\truetheta, \truepsi) \le \alpha\} ] = \alpha.\nn
 \end{align}
The first equality follows by definition of $H_0$ in~\eqref{eq:H02}; the second follows from the indicator identity
  $\Ind(a \le b) \le \Ind(c \le b)$ when $a\ge c$; 
  and the third follows from Theorem~\ref{thm1}. Since $\truetheta\in\Theta$ w.p.~1 it follows that $\pr(\truetheta\in\Theta_{1-\alpha}) \ge 1-\alpha$.
\end{proof}

\newcommand{\PV}{\pval(\truetheta, \truepsi)}
\newcommand{\numeq}[1]{\quad\quad~\mathrm{(#1)}}
\section*{Proof of Theorem~\ref{thm3}}
\begin{theorem}\label{thm3}
\ThmThree
\end{theorem}
\begin{proof}
Following a similar reasoning as above:
\begin{align}
\pr(\theta_0\notin\hat\Theta_{1-\alpha} \mid H_0)
 & = \pr\{ \pval(\truetheta, \hat\psi_{\truetheta}) \le \alpha \mid H_0 \} \numeq{i}
 \nonumber\\
 & = \pr\{\PV +  \underbrace{\pval(\truetheta, \hat\psi_{\truetheta}) - \PV}_{\Delta}  \le \alpha \mid H_0  \} \nonumber\\
& \le  \pr\{ \PV  \le  \alpha+  |\Delta| \mid H_0 \} \numeq{ii}
\nonumber\\
& \le \pr(|\Delta | > \epsilon \mid H_0 ) + \pr\{ \PV \le \alpha + \epsilon \mid H_0 \}
\numeq{iii}
\nonumber\\
& = \alpha + \epsilon + P(|\Delta | > \epsilon \mid H_0 ). 
\numeq{iv}
\nonumber\\
& \le \alpha + \min_\epsilon\{\epsilon + P(|\Delta | > \epsilon\mid H_0 )\}
\numeq{v}
\nn\\
& = \alpha + o_P(1).\nn
\numeq{vi}
\end{align}
In these derivations, line (i) follows from definition of the confidence set and $H_0$; 
line (ii) follows from the indicator identity $\Ind(A + B \le a) \le \Ind(A\le a + |B|)$ for any scalar random variables $A, B$ and $a\in\Real$; 
line (iii) follows from the 
indicator identity $\Ind(A \le a + B) \le \Ind(B > \epsilon) + \Ind(A \le a + \epsilon)$ for any $\epsilon <0$, 
and $A, B, a$ as above; line (iv) follows from Theorem~\ref{thm1}; 
line (v) gives the tightest bound exploiting the fact that  $\epsilon>0$ is arbitrary.
Finally, line (vi) follows from consistency of $\hat\psi_{\truetheta}$.
\end{proof}

\subsection*{Proof of Theorem~\ref{thm:np}}
\begin{theorem}\label{thm:np}
\ThmFour
\end{theorem}
\begin{proof}
The proof follows the same steps as Theorem~\ref{thm1} 
as we only have to derive that $s_n \eqd \sobs \mid T_n$.
The only difference is that this equivalence is implied 
immediately from the assumed invariance based on $\Pi(\Tn; \theta)$ rather than relying 
on the correct periodic component $\yp$ as in Theorem~\ref{thm1}.
\end{proof}

\end{document}